\documentclass[journal]{IEEEtran}
\IEEEoverridecommandlockouts
\usepackage{lineno}
\usepackage{hyperref}
\usepackage{cite}
\usepackage{amsmath,amssymb,amsfonts,cases} 
\usepackage{amsmath}
\usepackage{amsthm}
\DeclareMathOperator*{\argmax}{argmax}
\DeclareMathOperator*{\argmin}{argmin}
\usepackage{graphicx}
\usepackage{textcomp}
\usepackage{xcolor}
\usepackage{graphicx}
\usepackage{float}
\usepackage{subfigure}
\usepackage{amsmath,mleftright}
\usepackage{amsfonts,amssymb}
\usepackage{mathrsfs}
\usepackage{mathtools}
\usepackage{algorithm}
\usepackage{algorithmic}
\usepackage{bm}
\usepackage{multirow}
\usepackage{array}
\usepackage{amssymb}
\usepackage{amsmath}
\usepackage{cite}
\usepackage{url}
\usepackage{xcolor}
\usepackage{cite,graphicx,amsmath,amssymb}
\usepackage{subfigure}
\usepackage{fancyhdr}
\usepackage{mdwmath}
\usepackage{mdwtab}
\usepackage{caption}
\usepackage{amsthm}
\usepackage{setspace}
\usepackage{bm}
\usepackage{mathtools}
\usepackage{dsfont}
\usepackage{bbm}
\usepackage{framed}
\newtheorem{remark}{Remark}
\newtheorem{theorem}{Theorem}

\newtheorem{lemma}{Lemma}

\newtheorem{corollary}{Corollary}

\makeatletter
\newcommand{\biggg}{\bBigg@{3}}
\newcommand{\Biggg}{\bBigg@{3.5}}
\makeatother
\makeatletter
\renewcommand{\maketag@@@}[1]{\hbox{\m@th\normalsize\normalfont#1}}%
\makeatother
\def\BibTeX{{\rm B\kern-.05em{\sc i\kern-.025em b}\kern-.08em
    T\kern-.1667em\lower.7ex\hbox{E}\kern-.125emX}}
    \expandafter\def\expandafter\normalsize\expandafter{%
    \normalsize%
    \setlength\abovedisplayskip{4pt}%
    \setlength\belowdisplayskip{4pt}%
    \setlength\abovedisplayshortskip{2pt}%
    \setlength\belowdisplayshortskip{2pt}%
}
\allowdisplaybreaks[4]

\begin{document}
\title{Spatial Reuse in Multi-Cell Pinching-Antenna Systems}
\author{Zhenqiao Cheng, Chongjun~Ouyang, and Zhiguo~Ding,~\IEEEmembership{Fellow,~IEEE}\vspace{-10pt}}
\maketitle

\begin{abstract}
In this paper, spatial reuse in a multi-cell pinching-antenna system (PASS) is investigated, where each cell employs a dielectric waveguide with one activated pinching antenna (PA) to serve a single user. Spatial reuse refers to the simultaneous use of the same time-frequency resource block (RB) by multiple spatially separated cells. Due to inter-cell interference (ICI), simultaneous transmission by all cells on the same RB may be infeasible, and multiple RBs may be required to serve all users. The spatial-reuse capability is thus characterized by the minimum number of RBs required to satisfy all users' signal-to-interference-plus-noise ratio (SINR) targets under per-cell power budgets. Flexible PA placement can enhance the desired-link gain of each user and reduce the required transmit power of its serving cell, which in turn mitigates ICI and allows more cells to share the same RB. To exploit these benefits, a joint optimization problem of RB allocation, PA placement, and power control is formulated. For the two-cell case, the globally optimal PA placement is determined by minimizing the peak transmit power required by the two cells. A one-dimensional special case is further analyzed to derive analytical expressions for the average RB requirement and its scaling laws with respect to cell length under different PA-placement strategies. For the general multi-cell case, the optimal two-cell solution is exploited to construct a weighted compatibility graph, whose edge weights represent the minimum peak transmit powers required for feasible pairwise RB reuse. Based on this graph, a greedy RB allocation algorithm with minimum-weight pair initialization and element-wise PA-placement optimization is developed. Numerical results validate the analytical findings and demonstrate that PASS significantly reduces the required RBs compared with conventional fixed-antenna systems.
\end{abstract}
\begin{IEEEkeywords}
Multi-cell, pinching antenna, resource block, spatial reuse.
\end{IEEEkeywords}
\section{Introduction}
\label{sec:introduction}

Reconfigurable antennas (RAs) have attracted growing attention due to their ability to adapt wireless channels to different quality-of-service (QoS) requirements. Representative examples include fluid/movable antennas \cite{zhu2024movable,wong2020fluid}, reconfigurable intelligent surfaces \cite{wu2019towards}, and dynamic metasurface antennas \cite{shlezinger2021dynamic}. By adjusting antenna positions or electromagnetic responses, these technologies provide additional degrees of freedom (DoFs) for wireless transmission. However, their channel reconfigurability is generally constrained within a wavelength-scale region. Such spatial reconfigurability may be insufficient to overcome severe large-scale path loss and signal blockage, particularly at higher carrier frequencies, such as the upper mid-band spectrum considered for future wireless networks.

To address these limitations, NTT DOCOMO introduced the concept of pinching-antenna system (PASS) \cite{suzuki2022pinching}. In PASS, radio-frequency signals propagate through a dielectric waveguide, while dielectric pinching antennas (PAs) attached to the waveguide couple energy between the guided wave and free space \cite{liu2025pinching}. The locations of these PAs can be flexibly adjusted along the waveguide. Since a waveguide can extend over tens or even hundreds of meters, PASS offers a much larger spatial reconfiguration range than conventional RAs. In particular, a PA can be deployed close to its intended user to establish a favorable line-of-sight (LoS) link and substantially alleviate large-scale path loss and signal blockage \cite{suzuki2022pinching,liu2025pinching,ding2024flexible}.

These attractive properties have stimulated extensive research on PASS, as summarized in recent surveys \cite{liu2025pinchingtutorial,lu2026survey,liu2025pinchingsurvey}. Most existing studies have focused on single-cell systems and investigated physical-layer transmission designs, such as PA placement, beamforming, and power allocation. More recently, several studies have explored the application of PASS to multi-cell networks. For example, stochastic-geometry-based analyses have been developed to characterize network throughput \cite{zhu2025topological,sun2025stochastic}, while other studies have investigated the joint optimization of beamforming, PA placement, and base station (BS)-user association \cite{chen2026pinching,li2026pinching,jiang2026pinching}. Despite these efforts, the existing investigations have mainly concentrated on the transmission-performance gains offered by PASS, while its fundamental implications for network-level resource management remain less understood.

In this regard, recent studies have revealed the potential of flexible PA placement for inter-cell interference (ICI) management in multi-cell PASS \cite{ding2026toward,ding2026characterizing}. Specifically, by placing a PA close to its intended user, the desired-link gain can be improved and the transmit power required to satisfy the QoS target can be reduced \cite{ding2026toward}. This reduction in transmit power can further alleviate the interference caused to neighboring cells \cite{ding2026characterizing}. Moreover, flexible PA placement can directly reshape the interference channels between cells and provide additional opportunities for ICI mitigation. These findings indicate that PA placement not only improves individual communication links but also provides a new DoF for interference management at the network level.

Motivated by these observations, this paper investigates another important network-level benefit of PASS, namely \emph{spatial reuse}. In multi-cell networks, spatial reuse refers to the simultaneous use of the same time-frequency resource block (RB) by multiple spatially separated cells. Such simultaneous transmission is feasible only if the QoS requirements of all involved users can be satisfied within the transmit-power budget of each cell. In networks equipped with fixed antennas or conventional aperture-limited RAs, serving a cell-edge user may still require high transmit power and cause severe ICI. Consequently, simultaneous transmission by all cells may become infeasible, and different groups of cells must be scheduled on separate RBs. 

In contrast, PASS allows dielectric waveguides to extend across individual cells, particularly in small-cell deployments. Flexible PA placement can reduce the required transmit power and mitigate ICI \cite{ding2026toward,ding2026characterizing}. These capabilities enable more cells to share the same RB and potentially reduce the total RB requirement. This motivates the investigation of spatial reuse in multi-cell PASS through the joint design of PA placement, power control, and RB allocation. The main contributions of this paper are summarized as follows.
\begin{itemize}
    \item We investigate spatial reuse in multi-cell PASS, where each cell employs a dielectric waveguide with one activated PA to serve a single user. The waveguides extend across their respective cells to provide flexible PA placement and improve user coverage. We characterize the network-wide spatial-reuse capability through the minimum number of RBs required to satisfy all users' signal-to-interference-plus-noise ratio (SINR) requirements under the individual transmit-power budget of each cell. To this end, we formulate a joint optimization problem of RB allocation, PA placement, and power control.

    \item We first investigate the two-cell case and transform the RB-reuse feasibility problem into the minimization of the maximum required transmit power through PA placement. By exploiting the Karush-Kuhn-Tucker (KKT) conditions and a polynomial eigenvalue formulation, we characterize the interior stationary points and boundary candidates to determine the globally optimal PA placement. To obtain further insights, we investigate a one-dimensional (1D) two-cell case and characterize the average RB requirements under nearest-PA, feasibility-optimal, and power-optimal PA placement. We further derive their asymptotic scaling laws as the cell length increases and establish a strictly positive asymptotic RB gap between PASS and the conventional fixed-antenna system.
    
    \item We extend the investigation to general multi-cell networks and develop a low-complexity algorithm for joint RB allocation, PA placement, and power control. Specifically, we exploit the optimal two-cell solution to construct a weighted compatibility graph, where each node represents a cell and an edge connects two cells that can share the same RB under their respective QoS and transmit-power constraints. The weight of each edge is defined as the minimum peak transmit power required for simultaneous transmission by the corresponding two cells. Based on this graph, we progressively construct reuse groups by selecting minimum-weight feasible pairs and admitting additional cells through PA-placement refinement and power-feasibility checks. To this end, we develop an element-wise alternating optimization algorithm to refine the PA positions within each reuse group.

    \item We present numerical results to validate the theoretical analysis and evaluate the proposed designs. The results demonstrate that: i) PASS significantly reduces the average RB requirement compared with conventional fixed-antenna systems in both two-cell and multi-cell networks; ii) in the two-cell case, power-optimal PA placement achieves the lowest RB requirement, followed by feasibility-optimal and nearest-PA placement, although their performance differences are small; and iii) the proposed multi-cell algorithm significantly reduces the required RBs and improves the one-RB reuse probability, while nearest-PA placement combined with RB allocation achieves performance close to the proposed joint design.

\end{itemize}

The remainder of this paper is organized as follows. Section~\ref{sec:system} introduces the multi-cell PASS model. Section~\ref{sec:two_cell} investigates the two-cell case, while Section~\ref{sec:multicell_design} extends the investigation to general multi-cell networks. Section~\ref{sec:numerical_results} presents numerical results, and Section~\ref{sec:conclusion} concludes the paper.

\noindent\emph{Notation:} Scalars, vectors, and matrices are denoted by nonbold, bold lowercase, and bold uppercase letters, respectively. For a matrix $\mathbf A$, $\mathbf A^{\mathsf T}$, $\mathbf A^{-1}$, and $[\mathbf A]_{i,j}$ denote its transpose, inverse, and $(i,j)$th element, respectively. Moreover, $[\mathbf a]_i$ denotes the $i$th element of vector $\mathbf a$, while $\mathbf I_N$ and $\mathbf 0$ denote the $N\times N$ identity matrix and a zero matrix of appropriate dimensions, respectively. The sets $\mathbb C$ and $\mathbb R$ denote the complex and real numbers, respectively. Finally, $\mathbb E[\cdot]$ denotes the expectation operator.

\begin{figure}[!t]
 \centering
\includegraphics[width=0.45\textwidth]{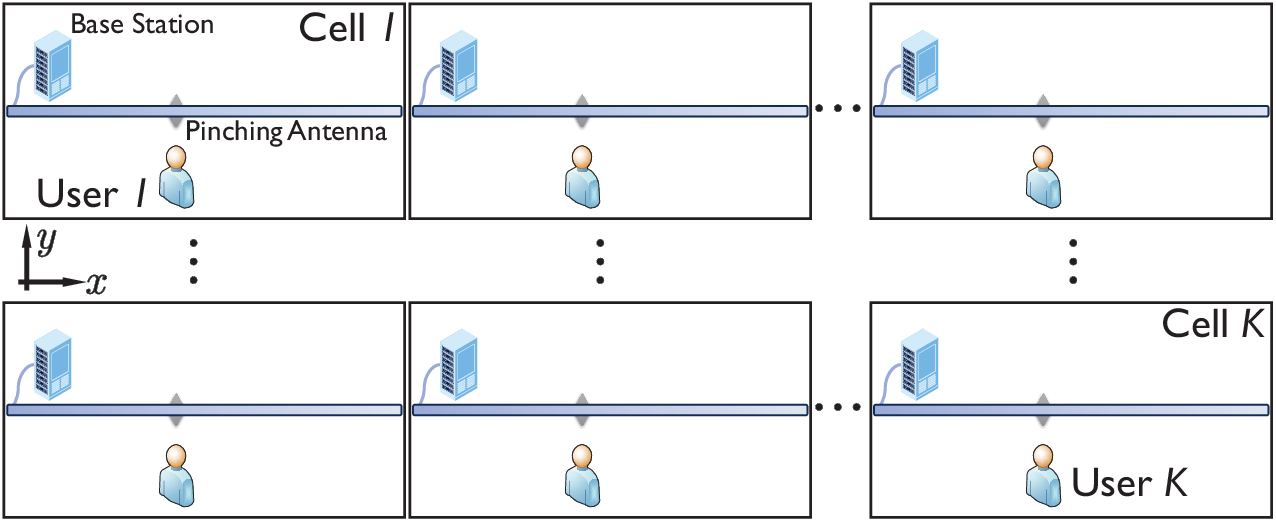}
\caption{Illustration of a multi-cell PASS with independent waveguides aligned along the $x$-direction.}
\label{Figure: System_Model}
\vspace{-10pt}
\end{figure}

\section{System Model}
\label{sec:system}
Consider the multi-cell PASS illustrated in {\figurename} {\ref{Figure: System_Model}}. The service region consists of $K$ adjacent rectangular cells with nonoverlapping interiors. Let $\mathcal{K}\triangleq\{1,\ldots,K\}$ denote the cell index set. Each cell has length $D_x$ along the $x$-axis and width $D_y$ along the $y$-axis. The center of cell $k$ is located at $(x_k^{\rm C},y_k^{\rm C})$, and its service region is defined as $
\mathcal{C}_k
\triangleq
[x_k^{\rm{C}}-\frac{D_x}{2},x_k^{\rm{C}}+\frac{D_x}{2}]
\times
[y_k^{\rm{C}}-\frac{D_y}{2},y_k^{\rm{C}}+\frac{D_y}{2}]$. Each cell is equipped with an independent dielectric waveguide that extends along the $x$-direction over the entire cell length at height $H>0$ above the user plane. The waveguide is aligned with the centerline of its cell, such that waveguides within the same row are collinear, while those in different rows are parallel. One PA is activated on each waveguide when the corresponding cell is scheduled. Let $s_k\in[-\frac{D_x}{2},\frac{D_x}{2}]\triangleq{\mathcal{S}}$ denote the PA position relative to the center of cell $k$. The corresponding three-dimensional location is $\mathbf{r}_k(s_k)=[x_k^{\rm C}+s_k,y_k^{\rm C},H]^{\mathsf T}$.

Each cell has one associated user to be served during a scheduling period. The location of user $k$ is denoted by $\mathbf{u}_k=[x_k^{\rm U},y_k^{\rm U},0]^{\mathsf T}$, where $(x_k^{\rm U},y_k^{\rm U})\in\mathcal{C}_k$. Each active cell transmits an independent data stream to its associated user without inter-cell data sharing or joint transmission. We assume that the user locations and position-dependent channel gains are perfectly known and that the propagation environment remains unchanged during each scheduling period.

\subsection{Channel Model}
\label{subsec:channel}

PASS is particularly attractive for high-frequency communications \cite{suzuki2022pinching}, where LoS propagation typically dominates \cite{ouyang2024primer}. We thus consider LoS propagation between the PAs and users. The free-space channel coefficient from the PA in cell $k'$ to user $k$ is modeled as follows:
\begin{align}
h_{k',k}^{\rm LoS}(s_{k'})
=\frac{\sqrt{\eta}\,{\rm e}^{-{\rm j}k_0\|\mathbf{r}_{k'}(s_{k'})-\mathbf{u}_k\|}}
{\|\mathbf{r}_{k'}(s_{k'})-\mathbf{u}_k\|},\nonumber
\end{align}
where $\eta\triangleq\frac{\lambda^2}{16\pi^2}$, $\lambda$ is the free-space wavelength, and $k_0\triangleq\frac{2\pi}{\lambda}$ is the free-space wavenumber. In addition to free-space propagation, the signal experiences phase shifts and attenuation during its propagation along the dielectric waveguide from the feed point to the activated PA. Let $\mathbf{r}_{k'}^{\rm Feed}$ denote the feed-point location of waveguide $k'$. The corresponding in-waveguide channel coefficient is modeled as follows:
\begin{align}
h_{k'}^{\rm In}(s_{k'})
={\rm e}^{-{\rm j}n_{\rm eff}k_0
\|\mathbf{r}_{k'}(s_{k'})-\mathbf{r}_{k'}^{\rm Feed}\|} 10^{-\frac{\kappa
\|\mathbf{r}_{k'}(s_{k'})-\mathbf{r}_{k'}^{\rm Feed}\|}{20}},
\nonumber
\end{align}
where $n_{\rm eff}$ is the effective refractive index of the dielectric waveguide, and $\kappa$ denotes the waveguide attenuation coefficient in dB/m \cite{yeh2008essence}.

The end-to-end channel coefficient from cell $k'$ to user $k$ is therefore given by $h_{k',k}(s_{k'})=h_{k'}^{\rm In}(s_{k'})h_{k',k}^{\rm LoS}(s_{k'})$. As theoretically demonstrated in \cite{xu2026pinching,xu2026pinching2}, the impact of in-waveguide attenuation on system performance and PA placement can be negligible under practical deployment conditions. Moreover, the segmented-waveguide architecture proposed in \cite{ouyang2025uplink} can further reduce such attenuation through shorter waveguide segments connected to the feed point via low-loss wired links. Motivated by these findings, we neglect in-waveguide attenuation in the subsequent analysis to facilitate the characterization of PA placement and spatial reuse. Accordingly, the channel power gain is approximated as follows:
\begin{align}
g_{k',k}(s_{k'})
&\triangleq |h_{k',k}(s_{k'})|^2
=
|h_{k'}^{\rm In}(s_{k'})h_{k',k}^{\rm LoS}(s_{k'})|^2\nonumber\\
&\approx
|h_{k',k}^{\rm LoS}(s_{k'})|^2=\frac{\eta}
{q_{{k'},k}+(s_{k'}-\xi_{{k'},k})^2},\label{eq:channel_power_gain}
\end{align}
where $q_{{k'},k}\triangleq (y_k^{\rm U}-y_{k'}^{\rm C})^2+ H^2>0$ and $\xi_{{k'},k}\triangleq x_k^{\rm U}-x_{k'}^{\rm C}$.

\subsection{Multi-Cell Transmission}
\label{subsec:multicell_transmission}

Consider a set $\mathcal{A}\subseteq\mathcal{K}$ of cells that transmit simultaneously over the same time-frequency resource. Let $p_k$ denote the transmit power of cell $k$, subject to $0\leq p_k\leq P_{\max}$. The received SINR at user $k\in\mathcal{A}$ is given by
\begin{align}
\Gamma_k
=
\frac{p_k g_{k,k}(s_k)}
{\sum_{k'\in\mathcal{A}\setminus\{k\}}
p_{k'}g_{k',k}(s_{k'})
+\sigma^2},
\nonumber
\end{align}
where $\sigma^2>0$ is the receiver noise power. Each user has a common SINR target $\gamma>0$. To satisfy $\Gamma_k\geq\gamma$, the transmit power of cell $k$ must satisfy
\begin{align}
p_k\geq
\frac{\gamma}
{g_{k,k}(s_k)}
\left(
\sum\nolimits_{k'\in\mathcal{A}\setminus\{k\}}
p_{k'}g_{k',k}(s_{k'})
+\sigma^2
\right).
\label{eq:power_inequality}
\end{align}
Equation~\eqref{eq:power_inequality} reveals the coupling among transmit power, PA placement, and ICI.

In networks equipped with conventional fixed or aperture-limited RAs, serving a cell-edge user may require high transmit power due to the large propagation distance. This can cause substantial ICI to users in neighboring cells and further increase the required power to satisfy their QoS targets. Therefore, simultaneous transmission by all cells may become infeasible under the per-cell power budget $P_{\max}$, and multiple orthogonal RBs may be required to serve all users.
\vspace{-5pt}
\begin{remark}
PASS provides additional spatial flexibility to alleviate this limitation. By placing the PA close to its intended user, each cell can improve its desired-link gain and reduce the transmit power required to satisfy the SINR target \cite{ding2026toward}. The reduced transmit power can alleviate ICI \cite{ding2026characterizing}, while PA placement can also reshape the interference channels between cells. Therefore, joint PA placement and power control can enhance spatial reuse and reduce the number of required RBs.
\end{remark}
\vspace{-5pt}
\subsection{Problem Formulation}
\label{subsec:rb_problem}

Our objective is to \emph{minimize the number of orthogonal RBs} required to serve all $K$ users. Cells assigned to the same RB transmit simultaneously, whereas cells assigned to different RBs do not interfere. Each user is scheduled on exactly one RB. For an arbitrary nonempty active set $\mathcal{A}\subseteq\mathcal{K}$, we define $Q^\star(\mathcal{A})$ as the minimum peak transmit power required to satisfy the SINR targets through joint PA placement and power control, i.e.,
\begin{subequations}\label{eq:group_problem_all}
\begin{align}
Q^\star(\mathcal{A})
\triangleq
\min_{\{p_k\geq 0,s_k\in\mathcal{S}\}_{k\in\mathcal{A}}}
\quad&
\max_{k\in\mathcal{A}}~p_k\\
\operatorname{s.t.}\quad&
\Gamma_k\geq\gamma,~\forall k\in\mathcal{A}.
\end{align}
\end{subequations}
We set $Q^\star(\mathcal{A})=\infty$ if the SINR constraints cannot be satisfied for any admissible PA configuration. Accordingly, the cells in $\mathcal{A}$ can reuse one RB if and only if $Q^\star(\mathcal{A})\leq P_{\max}$.

Let $\{\mathcal{A}_1,\ldots,\mathcal{A}_T\}$ denote a partition of $\mathcal{K}$ into $T$ nonempty reuse groups, with each group assigned to one RB. We assume that every user can be served individually, i.e., $Q^\star(\{k\})\leq P_{\max}$ for all $k\in\mathcal{K}$. Otherwise, the corresponding user cannot satisfy its SINR target even without ICI. The RB consumption minimization problem is formulated as follows:
\begin{subequations}\label{eq:main_problem_all}
\begin{align} 
\min_{T,\{\mathcal{A}_t\}_{t=1}^{T}}
\quad&
T
\label{eq:main_problem}\\
\operatorname{s.t.}\quad&
\bigcup\nolimits_{t=1}^{T}\mathcal{A}_t=\mathcal{K},
\label{eq:partition_union}\\
&
\mathcal{A}_t\cap\mathcal{A}_{\ell}
=\varnothing,\quad t\neq\ell,
\label{eq:partition_disjoint}\\
&
Q^\star(\mathcal{A}_t)\leq P_{\max},
\quad t=1,\ldots,T,
\label{eq:partition_feasibility}\\
&
\mathcal{A}_t\neq\varnothing,
\quad t=1,\ldots,T.
\label{eq:partition_antenna}
\end{align}
\end{subequations}
Problem~\eqref{eq:main_problem_all} couples PA placement and power control with combinatorial RB allocation. To characterize the impact of PA placement on spatial reuse, we first investigate the two-cell case and derive analytical results for the minimum required transmit power and the conditions for sharing a single RB. We then extend the investigation to general multi-cell networks.

\section{Two-Cell Case}
\label{sec:two_cell}

We first investigate the two-cell case, where cells $1$ and $2$ each serve one user. The two cells are not necessarily adjacent.

\subsection{Problem Reformulation}
\label{subsec:two_cell_problem}

To facilitate the subsequent analysis, we express the PA positions relative to the projections of their intended users onto the respective waveguide axes. Referring to \eqref{eq:channel_power_gain}, define $x\triangleq s_1-\xi_{1,1}$ and $y\triangleq s_2-\xi_{2,2}$, where $x\in[\ell_1,u_1]$ and $y\in[\ell_2,u_2]$, with $\ell_1\triangleq-\frac{D_x}{2}-\xi_{1,1}$, $u_1\triangleq\frac{D_x}{2}-\xi_{1,1}$, $\ell_2\triangleq-\frac{D_x}{2}-\xi_{2,2}$, and $u_2\triangleq\frac{D_x}{2}-\xi_{2,2}$. For notational convenience, define $a(x)\triangleq q_{1,1}+x^2$, $b(x)\triangleq q_{1,2}+(x-d_1)^2$, $c(y)\triangleq q_{2,1}+(y-d_2)^2$, and $e(y)\triangleq q_{2,2}+y^2$, where $d_1\triangleq\xi_{1,2}-\xi_{1,1}$ and $d_2\triangleq\xi_{2,1}-\xi_{2,2}$. Accordingly, the channel power gains are expressed as $g_{1,1}=\frac{\eta}{a}$, $g_{1,2}=\frac{\eta}{b}$, $g_{2,1}=\frac{\eta}{c}$, and $g_{2,2}=\frac{\eta}{e}$, where the arguments of $a$, $b$, $c$, and $e$ are omitted for brevity.

For fixed PA positions $(x,y)$, the SINR constraints in \eqref{eq:power_inequality} reduce to the following:
\begin{align*}
p_1\geq\frac{\gamma(p_2g_{2,1}+\sigma^2)}{g_{1,1}},\quad
p_2\geq\frac{\gamma(p_1g_{1,2}+\sigma^2)}{g_{2,2}}.
\end{align*}
When a finite nonnegative power solution exists, both inequalities hold with equality at the componentwise minimum-power solution. Hence,
\begin{align}
p_1=\gamma\frac{a}{c}p_2
+\frac{\gamma\sigma^2}{\eta}a,\quad
p_2=\gamma\frac{e}{b}p_1
+\frac{\gamma\sigma^2}{\eta}e.
\label{eq:two_power_linear}
\end{align}
Define $\Delta(x,y)\triangleq b(x)c(y)-\gamma^2a(x)e(y)$. Solving \eqref{eq:two_power_linear} yields the minimum required transmit powers
\begin{align}\nonumber
p_1(x,y)
=\frac{\gamma\sigma^2}{\eta}
\frac{ab(c+\gamma e)}{\Delta(x,y)},~
p_2(x,y)
=\frac{\gamma\sigma^2}{\eta}
\frac{ec(b+\gamma a)}{\Delta(x,y)}.
\end{align}
A finite nonnegative power solution exists if and only if
\begin{align}\label{eq:two-cell_feasibility}
\Delta(x,y)>0
\iff
\gamma^2\frac{a(x)e(y)}{b(x)c(y)}<1.
\end{align}
Therefore, the minimum peak transmit power required by the two cells is given by
\begin{align}
Q_{1,2}^{\star}
=
\min_{\substack{
x\in[\ell_1,u_1],\,
y\in[\ell_2,u_2]\\
\Delta(x,y)>0
}}
\max\{p_1(x,y),p_2(x,y)\},
\label{eq:two_cell_problem}
\end{align}
where $Q_{1,2}^{\star}=\infty$ if no feasible PA configuration satisfies $\Delta(x,y)>0$. Consequently, the two cells can reuse the same RB if and only if $Q_{1,2}^{\star}\leq P_{\max}$.

\subsection{Exact Algebraic Solution}
\label{subsec:two_cell_solution}

Problem~\eqref{eq:two_cell_problem} can be equivalently reformulated as follows:
\begin{subequations}
\label{eq:two_cell_epigraph}
\begin{align}
\min_{x,y,v}\quad &v\\
\mathrm{s.t.}\quad
&p_i(x,y)\leq v,\quad i=1,2,\\
&\ell_1\leq x\leq u_1,\quad
\ell_2\leq y\leq u_2,\quad
\Delta(x,y)>0.
\end{align}
\end{subequations}
To obtain the global solution, we characterize the stationary candidates in the \emph{interior} of the PA-position box $[\ell_1,u_1]\times[\ell_2,u_2]$ and on its \emph{boundary}. The optimal solution is then obtained by comparing the objective values of all feasible candidates.

\subsubsection{Interior Region}

Consider the interior region $\ell_1<x<u_1$ and $\ell_2<y<u_2$. The Lagrangian of \eqref{eq:two_cell_epigraph} is given by
\begin{align*}
\mathcal{L}
=v+\lambda_1(p_1-v)+\lambda_2(p_2-v),
\end{align*}
where $\lambda_1,\lambda_2\geq0$ are the Lagrange multipliers. Stationarity with respect to $v$ yields $\lambda_1+\lambda_2=1$. Together with complementary slackness, the stationary candidates can be classified into three cases: $p_1>p_2$, $p_2>p_1$, and $p_1=p_2$. For convenience, define $E(x,y)\triangleq ec(b+\gamma a)-ab(c+\gamma e)$. Since $p_1-p_2=-\frac{\gamma\sigma^2}{\eta}\frac{E(x,y)}{\Delta(x,y)}$, the three cases correspond to $E(x,y)<0$, $E(x,y)>0$, and $E(x,y)=0$, respectively.

\emph{Case I: $p_1>p_2$.}
In this case, $\lambda_1=1$ and $\lambda_2=0$. The stationarity conditions $\frac{\partial p_1}{\partial x}=0$ and $\frac{\partial p_1}{\partial y}=0$ reduce to
\begin{align*}
T_1(x;y)&\triangleq
cxb^2-\gamma^2e(x-d_1)a^2=0,\\
S_2(y)&\triangleq
-d_2y^2
+(q_{2,1}+d_2^2-q_{2,2})y
+q_{2,2}d_2=0.
\end{align*}
Here, $S_2(y)=0$ is a quadratic equation, whereas $T_1(x;y)=0$ is a polynomial equation of degree at most five in $x$ for fixed $y$. Thus, the stationary candidates can be obtained by successively solving these two univariate polynomial equations using standard polynomial-root routines, such as MATLAB's \texttt{roots}. Only real solutions satisfying the interior position constraints, $\Delta(x,y)>0$, and $E(x,y)<0$ are retained.

\emph{Case II: $p_2>p_1$.}
Similarly, $\lambda_1=0$ and $\lambda_2=1$. The stationarity conditions reduce to
\begin{align*}
S_1(x)&\triangleq
-d_1x^2
+(q_{1,2}+d_1^2-q_{1,1})x
+q_{1,1}d_1=0,\\
T_2(y;x)&\triangleq
byc^2-\gamma^2a(y-d_2)e^2=0.
\end{align*}
The stationary candidates can be obtained by first solving the quadratic equation $S_1(x)=0$ and then the polynomial equation $T_2(y;x)=0$. Only real solutions satisfying $\ell_1<x<u_1$, $\ell_2<y<u_2$, $\Delta(x,y)>0$, and $E(x,y)>0$ are retained.

\emph{Case III: $p_1=p_2$.}
In this case, $E(x,y)=0$, and the common transmit power can be written as follows:
\begin{align*}
p_1=p_2
=\frac{\gamma\sigma^2}{\eta\Psi(x,y)},
\end{align*}
where $\Psi(x,y)\triangleq\frac{1}{a}-\frac{\gamma}{c}$. Define $h(x,y)\triangleq\frac{E(x,y)}{abce}=\frac{1}{a}+\frac{\gamma}{b}-\frac{1}{e}-\frac{\gamma}{c}$. The minimum required power is attained by maximizing $\Psi(x,y)$ subject to $h(x,y)=0$ and $\Delta(x,y)>0$. To characterize the interior stationary candidates, consider the Lagrangian $\mathcal{L}_{\rm eq}=-\Psi(x,y)+\mu h(x,y)$, where $\mu$ is the Lagrange multiplier associated with $h(x,y)=0$. The first-order stationarity conditions yield $\frac{\partial\Psi}{\partial x}=\mu\frac{\partial h}{\partial x}$ and $\frac{\partial\Psi}{\partial y}=\mu\frac{\partial h}{\partial y}$. Eliminating $\mu$ gives
\begin{align*}
\frac{\partial\Psi}{\partial x}
\frac{\partial h}{\partial y}
-
\frac{\partial\Psi}{\partial y}
\frac{\partial h}{\partial x}
=0.
\end{align*}
After substitution and simplification, this condition reduces to
\begin{align*}
H(x,y)
\triangleq
xyb^2c^2
-\gamma^2(x-d_1)(y-d_2)a^2e^2
=0.
\end{align*}
Hence, all interior candidates in this case are contained in the following algebraic system:
\begin{align}
E(x,y)=0,\qquad H(x,y)=0.
\label{eq:EH_system}
\end{align}

To solve \eqref{eq:EH_system}, note that $E(x,y)$ and $H(x,y)$ are polynomials of degree four and at most five in $y$, respectively. They can be expressed as follows:
\begin{align}
E(x,y)=\sum\nolimits_{i=0}^{4}e_i(x)y^i,~
H(x,y)=\sum\nolimits_{i=0}^{5}h_i(x)y^i,
\label{eq:EH_poly}
\end{align}
where the coefficients $e_i(x)$ and $h_i(x)$ are polynomials in $x$. Particularly, the coefficients of $E(x,y)$ are
$e_4=b+\gamma a$,
$e_3=-2d_2(b+\gamma a)$,
$e_2=(q_{2,1}+q_{2,2}+d_2^2)(b+\gamma a)-(1+\gamma)ab$,
$e_1=2d_2[ab-q_{2,2}(b+\gamma a)]$, and
$e_0=q_{2,2}(q_{2,1}+d_2^2)(b+\gamma a)
-(q_{2,1}+d_2^2+\gamma q_{2,2})ab$. Similarly, the coefficients of $H(x,y)$ are
$h_5=xb^2-\gamma^2(x-d_1)a^2$,
$h_4=d_2[\gamma^2(x-d_1)a^2-4xb^2]$,
$h_3=2(q_{2,1}+3d_2^2)xb^2
-2\gamma^2q_{2,2}(x-d_1)a^2$,
$h_2=2d_2[\gamma^2q_{2,2}(x-d_1)a^2
-2(q_{2,1}+d_2^2)xb^2]$,
$h_1=(q_{2,1}+d_2^2)^2xb^2
-\gamma^2q_{2,2}^2(x-d_1)a^2$, and
$h_0=\gamma^2d_2q_{2,2}^2(x-d_1)a^2$,
where the dependence of $e_i$ and $h_i$ on $x$ is omitted for brevity. According to the resultant theory \cite{cox1998using}, the two polynomials in \eqref{eq:EH_poly} have a common root in $y$ only if their Sylvester matrix is singular. The corresponding Sylvester matrix $\mathbf{S}(x)\in\mathbb{R}^{9\times9}$ is given by
\begin{align*}
\mathbf{S}(x)=
\left[
\begin{smallmatrix}
e_4&e_3&e_2&e_1&e_0&0&0&0&0\\
0&e_4&e_3&e_2&e_1&e_0&0&0&0\\
0&0&e_4&e_3&e_2&e_1&e_0&0&0\\
0&0&0&e_4&e_3&e_2&e_1&e_0&0\\
0&0&0&0&e_4&e_3&e_2&e_1&e_0\\
h_5&h_4&h_3&h_2&h_1&h_0&0&0&0\\
0&h_5&h_4&h_3&h_2&h_1&h_0&0&0\\
0&0&h_5&h_4&h_3&h_2&h_1&h_0&0\\
0&0&0&h_5&h_4&h_3&h_2&h_1&h_0
\end{smallmatrix}
\right].
\end{align*}
Consequently, the candidate values of $x$ satisfy $\det\mathbf{S}(x)=0$. Since the entries of $\mathbf{S}(x)$ are polynomials in $x$ of degree at most five, we can express it as $\mathbf{S}(x)=\sum\nolimits_{i=0}^{5}x^i\mathbf{S}_i$, where $\{\mathbf{S}_i\in\mathbb{R}^{9\times9}\}_{i=0}^{5}$ are constant matrices. Thus, the candidate values of $x$ can be determined from the polynomial eigenvalue problem as follows:
\begin{align*}
\mathbf{S}(x)\mathbf{z}
=
\sum\nolimits_{i=0}^{5}x^i\mathbf{S}_i
\mathbf{z}
=\mathbf{0},\quad \mathbf{z}\neq\mathbf{0}.
\end{align*}
This fixed-dimensional problem can be solved using standard numerical routines, such as MATLAB's \texttt{polyeig}. For each real root $x\in(\ell_1,u_1)$, the corresponding values of $y$ are obtained from the quartic equation $E(x,y)=0$. Only real solutions satisfying $H(x,y)=0$, the interior position constraints, and $\Delta(x,y)>0$ are retained.

\subsubsection{Boundary Region}

We next consider the boundary of the PA-position box. For $x=x_{\rm{b}}\in\{\ell_1,u_1\}$, problem \eqref{eq:two_cell_epigraph} reduces to a univariate minimization over $y$. By the KKT conditions, its interior stationary candidates satisfy at least one of the following polynomial equations:
\begin{align*}
S_2(y)=0,\quad T_2(y;x_{\rm{b}})=0,\quad E(x_{\rm{b}},y)=0.
\end{align*}
For $y=y_{\rm{b}}\in\{\ell_2,u_2\}$, the corresponding candidates satisfy
\begin{align*}
T_1(x;y_{\rm{b}})=0,\quad S_1(x)=0,\quad E(x,y_{\rm{b}})=0.
\end{align*}
The four corners of the PA-position box are also included. All candidates must satisfy the position constraints and $\Delta(x,y)>0$. Therefore, the boundary candidates can also be obtained through univariate polynomial-root routines.

\subsubsection{Global Solution}

Let $\mathcal{V}_1$, $\mathcal{V}_2$, and $\mathcal{V}_{\rm eq}$ denote the feasible interior candidates from the three cases, respectively, and let $\mathcal{V}_{\rm bd}$ denote the feasible boundary candidates. Define $\mathcal{V}\triangleq\mathcal{V}_1\cup\mathcal{V}_2\cup\mathcal{V}_{\rm eq}\cup\mathcal{V}_{\rm bd}$. After all feasible candidates have been identified, the global optimum is obtained as
\begin{align}
Q_{1,2}^{\star}
=
\min_{(x,y)\in\mathcal{V}}
\max\{p_1(x,y),p_2(x,y)\}.
\label{eq:global_pair_solution}
\end{align}
If no feasible configuration exists, $Q_{1,2}^{\star}=\infty$.

\subsection{A One-Dimensional Special Case}
\label{sec:two_cell_insight}

\begin{figure}[!t]
 \centering
\includegraphics[width=0.45\textwidth]{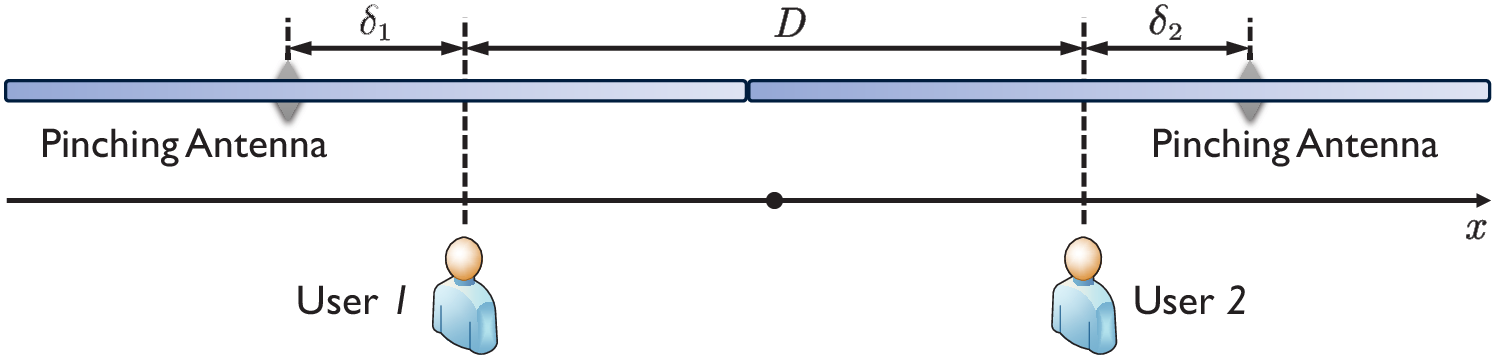}
\caption{Illustration of the 1D two-cell case.}
\label{Figure: Linear_Cell}
\vspace{-10pt}
\end{figure}

To obtain further analytical insights, we next focus on the 1D special case, where the waveguides and users are deployed along a line, as commonly encountered in vehicular and tunnel communication systems. Specifically, we consider two adjacent cells arranged along the $x$-axis, with horizontal extents $[-D_x,0]$ and $[0,D_x]$, as illustrated in Fig.~\ref{Figure: Linear_Cell}. The two waveguides are deployed at height $H$, and the users are located directly below their respective waveguides at $x_1^{\rm U}\in[-D_x,0]$ and $x_2^{\rm U}\in[0,D_x]$. Let $D\triangleq x_2^{\rm U}-x_1^{\rm U}>0$ denote the user separation. Without loss of generality, we shift the coordinate origin to the midpoint of the two users, whose locations become $\mathbf{u}_1=[-\frac{D}{2},0,0]^{\mathsf T}$ and $\mathbf{u}_2=[\frac{D}{2},0,0]^{\mathsf T}$. 

Since inward PA placement weakens the desired link and strengthens the interference link, it suffices to consider PA positions outward from the user projections (away from the common cell boundary). Let $\delta_1\in[0,x_1^{\rm U}+D_x]$ and $\delta_2\in[0,D_x-x_2^{\rm U}]$ denote the outward PA offsets from their respective user projections. The corresponding PA locations are given by $\mathbf{r}_1=[-\frac{D}{2}-\delta_1,0,H]^{\mathsf T}$ and $\mathbf{r}_2=[\frac{D}{2}+\delta_2,0,H]^{\mathsf T}$. For convenience, define the normalized user separation $d\triangleq \frac{D}{H}$ and the normalized outward PA offsets $x_i\triangleq\frac{\delta_i}{H}$ for $i\in\{1,2\}$. Let $A(x)\triangleq1+x^2$, $B(x)\triangleq1+(d+x)^2$, and $r(x)\triangleq\frac{A(x)}{B(x)}=\frac{1+x^2}{1+(d+x)^2}$, where $r(x)$ represents the cross-link-to-desired-link channel-gain ratio. With $p_0\triangleq\frac{\gamma\sigma^2H^2}{\eta}$, the minimum required transmit powers satisfy
\begin{align}\label{eq:two_powers_norm_general}
\frac{p_1}{p_0}
=
\frac{
A(x_1)\left[1+\gamma r(x_2)\right]
}{
1-\gamma^2r(x_1)r(x_2)
},~
\frac{p_2}{p_0}
=
\frac{
A(x_2)\left[1+\gamma r(x_1)\right]
}{
1-\gamma^2r(x_1)r(x_2)
}.
\end{align}
A finite nonnegative power solution exists if and only if $\gamma^2r(x_1)r(x_2)<1$.

\subsubsection{Nearest-PA Placement}
\label{subsec:nearest_positioning}

As a benchmark, consider nearest-PA placement, where each PA is positioned directly above its associated user, i.e., $x_1=x_2=0\triangleq x_{\rm{N}}$. This placement maximizes the desired-link gains without accounting for ICI. From \eqref{eq:two_powers_norm_general}, the two cells require the same transmit power:
\begin{align}
p_1=p_2
=p_0\frac{1+d^2}{1+d^2-\gamma}
\triangleq p_{\rm N}.
\label{eq:PN}
\end{align}
A finite-power solution exists if and only if $1+d^2>\gamma$. Thus, for $\gamma>1$, nearest-PA placement requires $d>d_{\rm F,N}\triangleq\sqrt{\gamma-1}$, whereas finite-power feasibility is guaranteed for any $d>0$ when $\gamma\leq1$. To account for the power budget, define $\rho\triangleq \frac{P_{\max}}{p_0}$. Since $p_{\rm N}>p_0$, one-RB reuse is impossible when $\rho\leq1$. We therefore assume $\rho>1$ in the sequel. Under this condition, $p_{\rm N}\leq P_{\max}$ is equivalent to $d^2\geq\gamma-1+\frac{\gamma}{\rho-1}$, which yields the following minimum normalized user separation required for one-RB reuse:
\begin{align}
d_{\rm N}(\rho)
=
\sqrt{
\max\left\{
0,\gamma-1+\frac{\gamma}{\rho-1}
\right\}
}.
\label{eq:dN_rho}
\end{align}
Therefore, nearest-PA placement supports one-RB reuse if and only if $d\geq d_{\rm N}(\rho)$. Despite maximizing the desired-link gains, nearest-PA placement may fail to support simultaneous transmission due to ICI.

\subsubsection{Feasibility-Optimal PA Placement}
\label{subsec:feasibility_optimal}

From \eqref{eq:two_powers_norm_general}, finite-power feasibility requires $\gamma^2r(x_1)r(x_2)<1$. Since $r(x)>0$, the feasibility margin is maximized when each PA independently minimizes $r(x)$. Differentiating $r(x)$ yields $r'(x)=\frac{2d(x^2+dx-1)}{[1+(d+x)^2]^2}$. Thus, $r(x)$ decreases for $0\leq x<x_{\rm I}$ and increases for $x>x_{\rm I}$, where $x_{\rm I}=\frac{\sqrt{d^2+4}-d}{2}$ is the unique positive root of $x^2+dx-1=0$.

Let $\bar{x}_i$ denote the maximum allowable normalized outward offset of PA $i$, where $\bar{x}_1=\frac{x_1^{\rm U}+D_x}{H}$ and $\bar{x}_2=\frac{D_x-x_2^{\rm U}}{H}$. The feasibility-optimal outward offset is therefore $\min\{x_{\rm I},\bar{x}_i\}$ for $i\in\{1,2\}$. For the following analysis, we assume $x_{\rm I}\leq\min\{\bar{x}_1,\bar{x}_2\}$, such that the feasibility-optimal offset is $x_1=x_2=x_{\rm I}$. Since $x_{\rm I}^2+dx_{\rm I}=1$, we have $d+x_{\rm I}=\frac{1}{x_{\rm I}}$ and hence $r(x_{\rm I})=x_{\rm I}^2$. The minimum value of $\gamma^2r(x_1)r(x_2)$ is therefore $\gamma^2x_{\rm I}^4$. Consequently, finite-power simultaneous transmission is feasible if and only if $\gamma x_{\rm I}^2<1$, or equivalently, $x_{\rm I}<\frac{1}{\sqrt{\gamma}}$. For $\gamma\leq1$, this condition always holds since $0<x_{\rm I}<1$ for any $d>0$. For $\gamma>1$, using $d=\frac{1}{x_{\rm I}}-x_{\rm I}$ yields the following user-separation condition:
\begin{align*}
d>\frac{\gamma-1}{\sqrt{\gamma}}\triangleq d_{\rm F}.
\end{align*}
Thus, no PA placement can support finite-power simultaneous transmission when $d\leq d_{\rm F}$, whereas $d>d_{\rm F}$ guarantees feasibility under the assumed PA-position condition. Moreover, $d_{\rm F}<d_{\rm F,N}=\sqrt{\gamma-1}$, which demonstrates that PA placement enlarges the feasible spatial-reuse region compared with nearest-PA placement.

Although $x_{\rm I}$ minimizes $r(x)$, it does not necessarily minimize the required transmit power because outward PA offset also reduces the desired-link gain. For $x_1=x_2=x_{\rm I}$, the required transmit power is given by
\begin{align}
p_1=p_2
=p_0\frac{1+x_{\rm I}^2}
{1-\gamma x_{\rm I}^2}
\triangleq p_{\rm I}.
\label{eq:PI}
\end{align}
Under the power budget $P_{\max}=\rho p_0$ with $\rho>1$, the condition $p_{\rm I}\leq P_{\max}$ is equivalent to $x_{\rm I}^2\leq\frac{\rho-1}{\rho\gamma+1}$. Using $d=\frac{1}{x_{\rm I}}-x_{\rm I}$, we obtain the minimum separation required for one-RB reuse:
\begin{align}
d_{\rm I}(\rho)
=
\sqrt{\frac{\rho\gamma+1}{\rho-1}}
-
\sqrt{\frac{\rho-1}{\rho\gamma+1}}.
\label{eq:dI_rho}
\end{align}
Therefore, feasibility-optimal PA placement supports one-RB reuse if and only if $d\geq d_{\rm I}(\rho)$.

\subsubsection{Power-Optimal PA Placement}
\label{subsec:power_optimal}

We next investigate the power-optimal PA placement. Following Section~\ref{subsec:feasibility_optimal}, we assume $x_{\rm I}\leq\min\{\bar{x}_1,\bar{x}_2\}$ and focus on $\gamma>1$ and $d>d_{\rm F}$. The cases with more restrictive PA-position constraints can be addressed using the exact algebraic solution in Section~\ref{subsec:two_cell_solution}.
\vspace{-5pt}
\begin{lemma}
\label{prop:symmetric_power_optimum}
For the considered two-cell case, the globally power-optimal PA placement satisfies $x_1^\star=x_2^\star$. Consequently, the corresponding transmit powers satisfy $p_1^\star=p_2^\star$.
\end{lemma}
\vspace{-5pt}
\begin{IEEEproof}
See Appendix~\ref{app:symmetric_power}.
\end{IEEEproof}
According to Lemma~\ref{prop:symmetric_power_optimum}, let $x_1=x_2=x$. The two cells require the same transmit power:
\begin{align*}
p_1=p_2
=\frac{p_0A(x)[1+\gamma r(x)]}
{1-\gamma^2r^2(x)}
=\frac{p_0A(x)}{1-\gamma r(x)}
=\frac{p_0}{J(x,d)},
\end{align*}
where $J(x,d)\triangleq\frac{1-\gamma r(x)}{A(x)}=
\frac{1}{1+x^2}
-\frac{\gamma}{1+(d+x)^2}$. Thus, minimizing the required transmit power is equivalent to maximizing $J(x,d)$. The power-optimal outward offset is given by
\begin{align}\label{eq:xP_def}
x_{\rm P}(d)
=
\argmax\nolimits_{0\leq x\leq x_{\rm I},1-\gamma r(x)>0}J(x,d),
\end{align}
and the corresponding minimum required transmit power is $p_{\rm P}^\star(d)=\frac{p_0}{\overline{J}(d)}$, where $\overline{J}(d)\triangleq J(x_{\rm P}(d),d)$. The following lemma characterizes the optimal PA placement.
\vspace{-5pt}
\begin{lemma}
\label{lemma:monotone}
For $d>d_{\rm F}$, $J(x,d)$ strictly increases over $0\leq x<x_{\rm P}(d)$ and strictly decreases over $x_{\rm P}(d)<x\leq x_{\rm I}$. The unique maximizer $x_{\rm P}(d)\in(0,x_{\rm I})$ satisfies $\frac{\partial J(x,d)}{\partial x}=0$, $J(x_{\rm P}(d),d)>0$, and $1-\gamma r(x_{\rm P}(d))>0$.
\end{lemma}
\vspace{-5pt}
\begin{IEEEproof}
See Appendix~\ref{app:optimal_displacement}.
\end{IEEEproof}
By Lemma~\ref{lemma:monotone}, $x_{\rm P}(d)$ is the unique maximizer of $J(x,d)$ and can be obtained through scalar bisection applied to $\frac{\partial J(x,d)}{\partial x}=0$ over $(0,x_{\rm I})$. We next characterize the minimum user separation required for one-RB reuse. Under the power-optimal PA placement, simultaneous transmission satisfies the power constraint $p_{\rm P}^\star(d)\leq P_{\max}$ if and only if $\overline{J}(d)\geq\frac{1}{\rho}$. The following lemma gives the monotonicity and range of $\overline{J}(d)$.
\vspace{-5pt}
\begin{lemma}
\label{lemma_opt_power}
For $d>d_{\rm F}$, $\overline{J}(d)$ is continuous and strictly increasing in $d$, with $0<\overline{J}(d)<1$. Furthermore, $\lim_{d\downarrow d_{\rm F}}\overline{J}(d)=0$ and $\lim_{d\rightarrow\infty}\overline{J}(d)=1$.
\end{lemma}
\vspace{-5pt}
\begin{IEEEproof}
See Appendix~\ref{app:power_threshold}.
\end{IEEEproof}
Since $\overline{J}(d)<1$, one-RB reuse is impossible for $\rho\leq1$. For $\rho>1$, Lemma~\ref{lemma_opt_power} guarantees the existence of a unique separation threshold $d_{\rm P}(\rho)>d_{\rm F}$, which satisfies
\begin{align}
\overline{J}\left(d_{\rm P}(\rho)\right)
=\frac{1}{\rho}.
\label{eq:dP_implicit}
\end{align}
Therefore, power-optimal PA placement supports one-RB reuse if and only if $d\geq d_{\rm P}(\rho)$ under the assumed PA-position constraints. Although $d_{\rm P}(\rho)$ does not admit a convenient closed-form expression, it can be obtained through two nested scalar bisection searches. Specifically, the inner search determines $x_{\rm P}(d)$ from $\frac{\partial J}{\partial x}=0$, whereas the outer search determines $d_{\rm P}(\rho)$ from \eqref{eq:dP_implicit}.

Finally, we characterize the power-optimal PA placement for large normalized user separation $d$. As $d\to\infty$, the feasibility-optimal outward offset satisfies
\begin{align*}
x_{\rm I}
=\frac{\sqrt{d^2+4}-d}{2}
=\frac{2}{\sqrt{d^2+4}+d}
=\frac{1}{d}+O(d^{-3}).
\end{align*}
Since $0<x_{\rm P}<x_{\rm I}$, we have $x_{\rm P}=O(d^{-1})$. From the stationarity condition $\frac{\partial J(x_{\rm P},d)}{\partial x}=0$, we obtain $\frac{x_{\rm P}}{(1+x_{\rm P}^2)^2}
=
\frac{\gamma(d+x_{\rm P})}
{[1+(d+x_{\rm P})^2]^2}$. Since $x_{\rm P}=O(d^{-1})$, the two sides admit the following asymptotic expansions:
\begin{align*}
\frac{x_{\rm P}}{(1+x_{\rm P}^2)^2}
&=x_{\rm P}\left[1+O(d^{-2})\right],\\
\frac{\gamma(d+x_{\rm P})}
{[1+(d+x_{\rm P})^2]^2}
&=\frac{\gamma}{d^3}
\left[1+O(d^{-2})\right].
\end{align*}
Equating these expressions yields
\begin{align*}
x_{\rm P}(d)=
\frac{\gamma}{d^3}
\frac{1+O(d^{-2})}{1+O(d^{-2})}=\frac{\gamma[1+O(d^{-2})]}{d^3}=\frac{\gamma}{d^3}+O(d^{-5}),
\end{align*}
as $d\to\infty$. Since $\delta_{\rm P}=Hx_{\rm P}$ and $d=\frac{D}{H}$, the corresponding physical offset satisfies $\delta_{\rm P}
=\frac{\gamma H^4}{D^3}
+O(\frac{H^6}{D^5})$.
\vspace{-5pt}
\begin{remark}
The feasibility-optimal and power-optimal outward offsets decrease as $d^{-1}$ and $d^{-3}$, respectively. Thus, both placements approach nearest-PA placement as the user separation increases, but the power-optimal placement converges much faster. This indicates that the power-optimal strategy increasingly prioritizes desired-link enhancement as ICI weakens with increasing user separation.
\end{remark}
\vspace{-5pt}
Finally, we compare the three PA-placement strategies. Nearest-PA placement maximizes the desired-link gain, feasibility-optimal placement minimizes the cross-link-to-desired-link channel-gain ratio $r(x)$, and power-optimal placement minimizes the required peak transmit power. From Lemma~\ref{lemma:monotone}, their optimal outward offsets satisfy
\begin{align*}
0=x_{\rm N}<x_{\rm P}<x_{\rm I}.
\end{align*}
This ordering reveals the tradeoff between desired-link enhancement and ICI reduction. In particular, power-optimal placement balances these two effects and achieves a strictly lower required transmit power than either benchmark whenever their required powers are finite. Consequently, under the assumed PA-position conditions, the corresponding one-RB reuse thresholds satisfy
\begin{align}
d_{\rm F}
<
d_{\rm P}(\rho)
<
\min\{d_{\rm I}(\rho),d_{\rm N}(\rho)\},
\label{eq:reuse_threshold_order}
\end{align}
where $\rho>1$. The first inequality follows from $\lim_{d\downarrow d_{\rm F}}\overline{J}(d)=0$, whereas the second follows from the strict optimality of $x_{\rm P}$ and the monotonicity of $\overline{J}(d)$.

\subsection{Average Resource Requirement and Scaling Analysis}
\label{subsec:average_rb}

We next characterize the average number of required RBs. Assume that the two users are independently and uniformly distributed within their respective cells, i.e., $x_1^{\rm U}\sim\mathcal{U}[-D_x,0]$ and $x_2^{\rm U}\sim\mathcal{U}[0,D_x]$. The user separation $D=x_2^{\rm U}-x_1^{\rm U}$ therefore follows a triangular distribution whose cumulative distribution function (CDF) is given by
\begin{align}\label{eq:D_cdf}
F_D(z)=
\begin{cases}
0, &z<0,\\[1mm]
\dfrac{z^2}{2D_x^2},
&0\leq z\leq D_x,\\[2mm]
1-\dfrac{(2D_x-z)^2}{2D_x^2},
&D_x<z\leq2D_x,\\[2mm]
1, &z>2D_x.
\end{cases}
\end{align}
If the required PA placements are attainable, the two cells need one RB when $D\geq Hd_s(\rho)$ for $s\in\{{\rm{N}},{\rm{I}},{\rm{P}}\}$ and two RBs otherwise. Intuitively, the average number of required RBs can be calculated as follows:
\begin{align}
\mathbb{E}[T_s]
=1+\Pr(D<Hd_s(\rho))
=1+F_D(Hd_s(\rho)).
\label{eq:ET_reference}
\end{align}
This expression is exact for nearest-PA placement. For feasibility-optimal and power-optimal placements, however, it serves as an approximation for finite cells, since the corresponding reuse thresholds were derived under the assumption $x_{\rm I}\leq\min\{\bar{x}_1,\bar{x}_2\}$, which may not hold for all user locations.

To obtain further analytical insights, we consider the large-cell regime with fixed $\gamma>1$ and normalized cell size $r\triangleq \frac{D_x}{H}\to\infty$. We scale the transmit-power budget according to the coverage requirement of a conventional fixed antenna located at the cell center. Specifically, the maximum distance from the antenna to a user is $\sqrt{H^2+D_x^2/4}$, and we set $P_{\max}=\frac{\gamma\sigma^2}{\eta}(H^2+\frac{D_x^2}{4})$. The corresponding normalized power budget is $\rho=\frac{P_{\max}}{p_0}=1+\frac{r^2}{4}$. For notational simplicity, we omit the dependence of $d_s(\rho)$ on $\rho$ in the sequel. The following lemma establishes that the average RB expression in \eqref{eq:ET_reference} becomes exact for sufficiently large cells.
\vspace{-5pt}
\begin{lemma}
\label{lemma:average_rb}
For fixed $\gamma>1$ and $\rho=1+\frac{r^2}{4}$, the average number of required RBs under the three PA-placement strategies satisfies
\begin{align}
\mathbb{E}[T_s]
=
1+\frac{H^2d_s^2}{2D_x^2}
=
1+\frac{d_s^2}{2r^2},
~ s\in\{{\rm N},{\rm I},{\rm P}\},
\label{eq:ET_large_cell}
\end{align}
for sufficiently large $r$.
\end{lemma}
\vspace{-5pt}
\begin{IEEEproof}
See Appendix~\ref{app:average_rb}.
\end{IEEEproof}
Based on Lemma~\ref{lemma:average_rb}, we next derive the asymptotic scaling laws for the three PA-placement strategies.
\subsubsection{Nearest-PA Placement}\label{sec:required_rb_Nearest-PA Placement}
Substituting $\rho
=1+\frac{r^2}{4}$ into \eqref{eq:dN_rho} yields
$d_{\rm N}^2=\gamma-1+\frac{4\gamma}{r^2}$.
Thus, $d_{\rm N}=O(1)$ as $r\to\infty$. From \eqref{eq:ET_large_cell}, the average number of required RBs satisfies
\begin{align}
\mathbb{E}[T_{\rm N}]
&=1+\frac{\gamma-1}{2r^2}
+\frac{2\gamma}{r^4}\nonumber\\
&=1+\frac{\gamma-1}{2}
\left(\frac{H}{D_x}\right)^2
+O\left(\frac{H^4}{D_x^4}\right).
\label{eq:ETN_asymp}
\end{align}

\subsubsection{Feasibility-Optimal PA Placement}\label{sec:required_rb_Feasibility-Optimal PA Placement}

Inserting $\rho
=1+\frac{r^2}{4}$ into \eqref{eq:dI_rho} gives
\begin{align*}
d_{\rm I}(r)
&=\frac{\sqrt{\gamma r^2+4(\gamma+1)}}{r}
-\frac{r}{\sqrt{\gamma r^2+4(\gamma+1)}}\\
&=\sqrt{\gamma}-\frac{1}{\sqrt{\gamma}}+O(r^{-2})=d_{\rm F}+O(r^{-2}),
\end{align*}
where $d_{\rm F}=\frac{\gamma-1}{\sqrt{\gamma}}$. Hence, $d_{\rm I}=O(1)$, and \eqref{eq:ET_large_cell} yields
\begin{align}
\mathbb{E}[T_{\rm I}]
&=1+\frac{d_{\rm F}^2}{2r^2}
+O(r^{-4})\nonumber\\
&=1+\frac{(\gamma-1)^2}{2\gamma}
\left(\frac{H}{D_x}\right)^2
+O\left(\frac{H^4}{D_x^4}\right).
\label{eq:ETI_asymp}
\end{align}

\subsubsection{Power-Optimal PA Placement}

From the reuse-threshold ordering in \eqref{eq:reuse_threshold_order}, we have $d_{\rm F}<d_{\rm P}(r)<d_{\rm I}(r)$. Since $d_{\rm I}(r)=d_{\rm F}+O(r^{-2})$, it follows that
$d_{\rm P}(r)=d_{\rm F}+O(r^{-2})$.
Therefore, the average number of required RBs satisfies
\begin{align}
\mathbb{E}[T_{\rm P}]
&=1+\frac{d_{\rm F}^2}{2r^2}
+O(r^{-4})\nonumber\\
&=1+\frac{(\gamma-1)^2}{2\gamma}
\left(\frac{H}{D_x}\right)^2
+O\left(\frac{H^4}{D_x^4}\right).
\label{eq:ETP_asymp}
\end{align}
\vspace{-10pt}
\begin{remark}\label{remark: convergence}
The above results show that all three PA-placement strategies asymptotically achieve one-RB reuse on average as $\frac{D_x}{H}\to\infty$. This behavior can be intuitively explained by the increasing user separation in larger cells. With PASS, each PA can remain close to its intended UE, which maintains a strong desired link while the ICI diminishes as the user separation increases. Consequently, the probability of requiring two RBs vanishes at a rate of $O(D_x^{-2})$. Notably, even nearest-PA placement achieves this scaling behavior, while feasibility-optimal and power-optimal placements further reduce the average RB requirement and attain the same leading-order scaling coefficient.
\end{remark}
\vspace{-5pt}
\subsubsection{Conventional Fixed-Antenna System}

For comparison, consider a conventional two-cell system with fixed antennas located at the cell centers, i.e., $-\frac{D_x}{2}$ and $\frac{D_x}{2}$, at height $H$. Define the normalized user coordinates $u=-\frac{x_1^{\rm U}}{D_x}$ and $v=\frac{x_2^{\rm U}}{D_x}$, where $u,v\sim\mathcal{U}[0,1]$. The normalized squared propagation distances are given by $a_{\rm C}\triangleq1+r^2(u-\tfrac12)^2$, $b_{\rm C}\triangleq1+r^2(v+\tfrac12)^2$, $c_{\rm C}\triangleq1+r^2(u+\tfrac12)^2$, and $e_{\rm C}\triangleq1+r^2(v-\tfrac12)^2$. Let $\Delta_{\rm C}\triangleq b_{\rm C}c_{\rm C}-\gamma^2a_{\rm C}e_{\rm C}$. From the two-cell power-control solution, the normalized minimum transmit powers are
\begin{align*}
\frac{p_{{\rm C},1}}{p_0}
=\frac{a_{\rm C}b_{\rm C}(c_{\rm C}+\gamma e_{\rm C})}
{\Delta_{\rm C}},\quad
\frac{p_{{\rm C},2}}{p_0}
=\frac{e_{\rm C}c_{\rm C}(b_{\rm C}+\gamma a_{\rm C})}
{\Delta_{\rm C}},
\end{align*}
provided that $\Delta_{\rm C}>0$; otherwise, finite-power simultaneous transmission is infeasible. Define $p_{\rm C}(u,v)=\max\{p_{{\rm C},1},p_{{\rm C},2}\}$ for $\Delta_{\rm C}>0$ and $p_{\rm C}(u,v)=\infty$ otherwise.

Unlike PASS, the conventional system has no separation-only reuse threshold, since its feasibility depends on both user locations. Although a simple closed-form expression for its average RB requirement is unavailable, the exact value can be obtained through numerical integration as follows:
\begin{align}\label{eq:ETC_analy}
\mathbb{E}[T_{\rm C}]
=
1+\int_0^1\int_0^1
\mathbbm{1}\left\{
p_{\rm C}(u,v)>P_{\max}
\right\}
\,{\rm d}u\,{\rm d}v,
\end{align}
where $\mathbbm{1}\{\cdot\}$ denotes the indicator function. 

We further characterize the asymptotic average RB requirement under the power normalization $\rho=\frac{P_{\max}}{p_0}=1+\frac{r^2}{4}$. As $r\to\infty$, the normalized minimum transmit powers satisfy
\begin{align*}
\frac{p_{{\rm C},1}}{P_{\max}}
\to f_\gamma(u,v),\quad
\frac{p_{{\rm C},2}}{P_{\max}}
\to f_\gamma(v,u),
\end{align*}
where
$f_\gamma(u,v)\triangleq
\frac{\alpha(u)+\alpha(v)\beta(u)}
{1-\beta(u)\beta(v)}$,
$\alpha(z)\triangleq4(z-\frac12)^2$, and
$\beta(z)\triangleq\gamma({z-\frac12})^2/({z+\frac12})^2$.
Define $\hat{f}_\gamma(u,v)=\max\{f_\gamma(u,v),f_\gamma(v,u)\}$ when $\beta(u)\beta(v)<1$, and $\hat{f}_\gamma(u,v)=\infty$ otherwise. Consequently, the asymptotic average RB requirement is given by
\begin{align}
\lim_{r\to\infty}\mathbb{E}[T_{\rm C}]
=
1+\int_0^1\int_0^1
\mathbbm{1}\{
\hat{f}_\gamma(u,v)>1
\}
\,{\rm d}u\,{\rm d}v.
\label{eq:ETC_asymp}
\end{align}
The limiting value depends only on $\gamma$ and can be evaluated numerically. To show that it is strictly greater than one, note that one-RB reuse requires $f_\gamma(u,v)\leq1$, which implies $\alpha(u)+\beta(u)\alpha(v)\leq1$, since $0<1-\beta(u)\beta(v)\leq1$ for feasible configurations. However, at $u=v=0$, we have $\alpha(0)=1$ and $\beta(0)=\gamma$, which yields $\alpha(0)+\beta(0)\alpha(0)=1+\gamma>1$ for any $\gamma>0$. By continuity, this necessary condition is also violated in a neighborhood of positive area around $(u,v)=(0,0)$, where $\mathbbm{1}\{\hat{f}_\gamma(u,v)>1\}=1$. Therefore, $\lim_{r\to\infty}\mathbb{E}[T_{\rm C}]>1$, which establishes a strictly positive excess RB requirement for conventional fixed antennas in the large-cell regime, whereas PASS asymptotically requires only one RB on average.

\subsubsection{Comparison and Asymptotic Insights}

We first compare the three PA-placement strategies. From \eqref{eq:ETI_asymp} and \eqref{eq:ETP_asymp}, we have
\begin{align*}
0\leq
\mathbb{E}[T_{\rm I}]-\mathbb{E}[T_{\rm P}]
=O(r^{-4}).
\end{align*}
Thus, feasibility-optimal placement achieves the same leading-order resource scaling as power-optimal placement, despite requiring higher transmit power in general. Moreover, \eqref{eq:ETN_asymp} and \eqref{eq:ETP_asymp} yield
\begin{align*}
\mathbb{E}[T_{\rm N}]-\mathbb{E}[T_{\rm P}]
=
\frac{\gamma-1}{2\gamma r^2}
+O(r^{-4}).
\end{align*}
Hence, nearest-PA placement also approaches one-RB reuse as $r\to\infty$, but with a larger leading-order coefficient.

We next compare PASS with the conventional fixed-antenna system. Since the conventional cell-center antenna positions are included in the feasible PA-placement regions, \emph{globally power-optimal PASS} cannot require more peak transmit power or RBs than the conventional fixed-antenna system. Moreover, the preceding analysis has established a strictly positive asymptotic RB gap between the conventional system and PASS. To further quantify this gap, we derive an explicit lower bound in terms of the SINR target $\gamma$. For the conventional system, finite-power simultaneous transmission requires $\gamma^2a_{\rm C}e_{\rm C}<b_{\rm C}c_{\rm C}$. However, for $0\leq u,v<\frac{1}{2}$,
\begin{align}
\lim_{r\to\infty}
\frac{\gamma^2a_{\rm C}e_{\rm C}}
{b_{\rm C}c_{\rm C}}
=
\gamma^2
\frac{(1-2u)^2(1-2v)^2}
{(1+2u)^2(1+2v)^2}.
\label{eq:conv_ratio_limit}
\end{align}
For $\gamma>1$, the right-hand side of \eqref{eq:conv_ratio_limit} exceeds one whenever $0\leq u,v<\epsilon_\gamma$, where $\epsilon_\gamma=\frac{\sqrt{\gamma}-1}{2(\sqrt{\gamma}+1)}$. Since this region has probability $\epsilon_\gamma^2$, it follows that
\begin{align*}
\liminf_{r\to\infty}\mathbb{E}[T_{\rm C}]
\geq(1-\epsilon_\gamma^2)+2\epsilon_\gamma^2=1+\epsilon_\gamma^2.
\end{align*}
In contrast, $\mathbb{E}[T_s]\to1$ for all three PA-placement strategies $s\in\{{\rm N},{\rm I},{\rm P}\}$. Consequently,
\begin{align*}
\liminf_{r\to\infty}
\left(
\mathbb{E}[T_{\rm C}]
-\mathbb{E}[T_{s}]
\right)
\geq\epsilon_\gamma^2=
\left(
\frac{\sqrt{\gamma}-1}
{2(\sqrt{\gamma}+1)}
\right)^2
>0.
\end{align*}
\vspace{-10pt}
\begin{remark}
The above results reveal a fundamental advantage of PASS over conventional fixed-antenna systems in spatial reuse. As cell coverage expands, conventional fixed antennas require higher transmit power to serve distant users, which can aggravate ICI and hinder RB reuse. Moreover, their fixed locations cannot adapt the relative strengths of desired and interference links, leaving a nonzero fraction of user configurations unable to share one RB even in the large-cell regime. In contrast, PASS can strengthen the desired links through flexible PA placement and thereby reduce the required transmit power and associated ICI. Meanwhile, PA placement can further reshape the cross-link gains to mitigate interference. These advantages facilitate simultaneous transmissions over the same RB and improve spectrum utilization. For larger networks, however, the aggregate interference from multiple cells must be jointly considered, which motivates the subsequent multi-cell design of RB allocation, PA placement, and power control.
\end{remark}
\vspace{-5pt}
\section{Multi-Cell Case}
\label{sec:multicell_design}

We now consider the multi-cell problem in \eqref{eq:main_problem_all}. The joint optimization of RB allocation, PA placement, and transmit power is a nonconvex mixed discrete-continuous problem, whose global solution generally requires exhaustive search over possible cell groupings. To reduce the computational complexity, we develop a power-weighted graph-assisted greedy RB allocation algorithm that exploits the optimal two-cell solution developed in Section~\ref{subsec:two_cell_solution}. Specifically, the algorithm initializes each reuse group using the minimum-weight feasible cell pair and progressively admits additional cells through PA-placement refinement and power-feasibility checks.

\subsection{Power-Weighted Compatibility Graph}
\label{subsec:pairwise_graph}

We first construct an undirected weighted graph $\mathcal{G}=(\mathcal{K},\mathcal{E})$ based on the optimal pairwise peak power $Q_{i,j}^{\star}=Q^{\star}(\{i,j\})$ derived in Section~\ref{subsec:two_cell_solution}. Specifically,
\begin{align*}
(i,j)\in\mathcal{E}
\quad\Longleftrightarrow\quad
Q_{i,j}^{\star}\leq P_{\max},
~
w_{i,j}=Q_{i,j}^{\star}.
\end{align*}
Thus, an edge connects cells $i$  and $j$ that can reuse the same RB, and its weight $w_{i,j}=Q_{i,j}^{\star}$ represents the minimum required peak transmit power. 
\vspace{-5pt}
\begin{remark}
A feasible reuse group must consist of pairwise compatible cells, i.e., every two cells in the group must be connected by an edge. However, pairwise compatibility alone does not guarantee multi-cell feasibility, since the aggregate ICI may violate the SINR constraints. Therefore, the graph is used to screen candidate cells, whereas the feasibility of an enlarged group is verified through joint PA placement and power control.
\end{remark}
\vspace{-5pt}

\subsection{Group-Wise PA-Placement Refinement}
\label{subsec:group_pa_placement}

For a given active set $\mathcal{A}=\{k_1,\ldots,k_{|\mathcal{A}|}\}$, we consider the group-wise joint PA-placement and power-control problem in \eqref{eq:group_problem_all}. We first optimize the transmit powers for fixed PA positions. Define $\mathbf{p}_{\mathcal{A}}\triangleq[p_{k_1},\ldots,p_{k_{|\mathcal{A}|}}]^{\mathsf T}$ and $\mathbf{s}_{\mathcal{A}}\triangleq[s_{k_1},\ldots,s_{k_{|\mathcal{A}|}}]^{\mathsf T}$. From \eqref{eq:power_inequality}, the componentwise minimum feasible power vector can be obtained by setting all SINR constraints to equality, i.e.,
\begin{align*}
p_k g_{k,k}(s_k)
-\gamma\sum\nolimits_{k'\in\mathcal{A}\setminus\{k\}}
p_{k'}g_{k',k}(s_{k'})
=\gamma\sigma^2,\quad k\in\mathcal{A}.
\end{align*}
These equations can be expressed in matrix form as follows:
\begin{align*}
[\mathbf{I}_{|\mathcal{A}|}-\mathbf{F}_{\mathcal{A}}(\mathbf{s}_{\mathcal{A}})]
\mathbf{p}_{\mathcal{A}}
=
\mathbf{b}_{\mathcal{A}}(\mathbf{s}_{\mathcal{A}}),
\end{align*}
where $\mathbf{b}_{\mathcal{A}}\in\mathbb{R}^{|\mathcal{A}|}$ and $\mathbf{F}_{\mathcal{A}}\in\mathbb{R}^{|\mathcal{A}|\times|\mathcal{A}|}$ have entries $[\mathbf{b}_{\mathcal{A}}]_m=\gamma\sigma^2/g_{k_m,k_m}(s_{k_m})$ and
\begin{align*}
[\mathbf{F}_{\mathcal{A}}]_{m,n}
=
\begin{cases}
\frac{\gamma g_{k_n,k_m}(s_{k_n})}
{g_{k_m,k_m}(s_{k_m})},
&m\neq n,\\
0,&m=n,
\end{cases}
\end{align*}
respectively, for $m,n\in\{1,\ldots,|\mathcal{A}|\}$. Provided that $\mathbf{I}_{|\mathcal{A}|}-\mathbf{F}_{\mathcal{A}}$ is invertible and the resulting power vector is positive, the minimum required transmit-power vector is given by
\begin{align}\label{eq:group_power_solution}
\mathbf{p}_{\mathcal{A}}^{\star}(\mathbf{s}_{\mathcal{A}})
=
[\mathbf{I}_{|\mathcal{A}|}-\mathbf{F}_{\mathcal{A}}(\mathbf{s}_{\mathcal{A}})]^{-1}
\mathbf{b}_{\mathcal{A}}(\mathbf{s}_{\mathcal{A}}).
\end{align}
Since this solution minimizes each power component, it also minimizes the required peak transmit power for fixed PA positions. By substituting \eqref{eq:group_power_solution} into \eqref{eq:group_problem_all}, the group-wise problem reduces to the following:
\begin{align*}
Q^{\star}(\mathcal{A})
=
\min\nolimits_{\mathbf{s}_{\mathcal{A}}\in\mathcal{S}^{|\mathcal{A}|}}
\Phi_{\mathcal{A}}(\mathbf{s}_{\mathcal{A}}),
\end{align*}
where $\Phi_{\mathcal{A}}(\mathbf{s}_{\mathcal{A}})
\triangleq
\|\mathbf{p}_{\mathcal{A}}^{\star}(\mathbf{s}_{\mathcal{A}})\|_{\infty}$, and placements that do not yield finite positive power solutions are assigned infinite cost. 

To address the coupling among the PA positions, we adopt an element-wise alternating optimization framework. Specifically, each PA position is optimized sequentially while the remaining PA positions are fixed. The corresponding subproblem is formulated as follows:
\begin{align*}
\min\nolimits_{s_k\in\mathcal{S}}
\Phi_{\mathcal{A}}(s_k,\mathbf{s}_{\mathcal{A}\setminus\{k\}}),\quad k\in{\mathcal{A}},
\end{align*}
where $\mathbf{s}_{\mathcal{A}\setminus\{k\}}$ contains the positions of the other PAs in the active set. Since this subproblem involves only one variable over a bounded interval, we employ a one-dimensional search. Specifically, the feasible interval $\mathcal{S}=[-\frac{D_x}{2},\frac{D_x}{2}]$ is discretized into a $Q$-point grid $\mathcal{Q}
=\{
-\frac{D_x}{2}
+\frac{(q-1)D_x}{Q-1}
\}_{q=1}^{Q}$. The PA position is then updated as follows:
\begin{align*}
s_k
\leftarrow
\argmin\nolimits_{s\in\mathcal{Q}\cup\{s_k\}}
\Phi_{\mathcal{A}}(s,\mathbf{s}_{\mathcal{A}\setminus\{k\}}),
\end{align*}
where candidate positions that do not yield finite positive power solutions are discarded. The optimization proceeds by sequentially updating all PA positions until the fractional decrease in the objective value falls below a predefined threshold or the maximum number of iterations is reached. Since the previous PA position is included in the search set, each update cannot increase the objective value.

Let $\widehat{\mathbf{s}}_{\mathcal{A}}$ denote the resulting PA placement and $\widehat{Q}(\mathcal{A})=\Phi_{\mathcal{A}}(\widehat{\mathbf{s}}_{\mathcal{A}})$ the corresponding required peak power. The active set is accepted as a feasible reuse group if $\widehat{Q}(\mathcal{A})\leq P_{\max}$.

\subsection{Stepwise RB Allocation}
\label{subsec:graph_resource_allocation}

We now combine the weighted compatibility graph with group-wise PA-placement refinement to construct the reuse groups. Specifically, we first select the minimum-weight edge among the unassigned cells and form an initial reuse group using the corresponding exact two-cell solution. We then progressively enlarge the group by examining unassigned cells that are connected to all its members.

For each candidate cell $m$, we temporarily form an enlarged group $\widetilde{\mathcal{A}}=\mathcal{A}\cup\{m\}$. To initialize its PA placement, we retain the positions of the existing group and perform a one-dimensional search over the newly added PA position $s_m$. If a finite positive power solution is found, we further refine all PA positions in $\widetilde{\mathcal{A}}$ using the element-wise optimization in Section~\ref{subsec:group_pa_placement}. The candidate is admitted if $\widehat{Q}(\widetilde{\mathcal{A}})\leq P_{\max}$. Otherwise, we retain the existing group and examine the next candidate. Once no additional cell can be admitted, the current group is assigned one RB. The procedure is repeated for the remaining cells until all cells are assigned. If no compatible pair remains, the remaining cells are assigned individual RBs. The overall procedure is summarized in Algorithm~\ref{alg:multicell_scheduling}.

\begin{algorithm}[t]
\caption{Greedy PA Placement and RB Allocation}
\label{alg:multicell_scheduling}
\footnotesize
\begin{algorithmic}[1]
\REQUIRE Cell set $\mathcal{K}$, compatibility graph $\mathcal{G}$, power limit $P_{\max}$.
\ENSURE Reuse groups $\{\mathcal{A}_t\}_{t=1}^{T}$, PA positions, and transmit powers.
\STATE Initialize $\mathcal{R}\leftarrow\mathcal{K}$ and $T\leftarrow0$.
\WHILE{$\mathcal{R}\neq\varnothing$}
    \IF{an edge exists among cells in $\mathcal{R}$}
        \STATE Select $(i,j)=\argmin_{(i,j)\in\mathcal{E},\,i,j\in\mathcal{R}}w_{i,j}$.
        \STATE Set $\mathcal{A}\leftarrow\{i,j\}$ using the exact two-cell solution.
    \ELSE
        \STATE Select $i\in\mathcal{R}$ and set $\mathcal{A}\leftarrow\{i\}$.
        \STATE Initialize $s_i$ at its nearest-PA position.
    \ENDIF
    \STATE $\mathcal{R}\leftarrow\mathcal{R}\setminus\mathcal{A}$.
    \STATE $\mathcal{C}\leftarrow\{m\in\mathcal{R}:(m,j)\in\mathcal{E},\ \forall j\in\mathcal{A}\}$.
    \WHILE{$\mathcal{C}\neq\varnothing$}
        \STATE Select $m\in\mathcal{C}$; $\mathcal{C}\leftarrow\mathcal{C}\setminus\{m\}$.
        \STATE Set $\widetilde{\mathcal{A}}\leftarrow\mathcal{A}\cup\{m\}$.
        \STATE Fix $\mathbf{s}_{\mathcal{A}}$ and search $s_m\in\mathcal{Q}$ for a finite positive power solution.
        \IF{a valid initial placement is found}
            \STATE Refine $\mathbf{s}_{\widetilde{\mathcal{A}}}$ by element-wise optimization.
            \IF{$\widehat{Q}(\widetilde{\mathcal{A}})\leq P_{\max}$}
                \STATE $\mathcal{A}\leftarrow\widetilde{\mathcal{A}}$; $\mathcal{R}\leftarrow\mathcal{R}\setminus\{m\}$.
                \STATE Retain the refined PA positions and powers.
                \STATE Update $\mathcal{C}\leftarrow\{j\in\mathcal{R}:(j,k)\in\mathcal{E},\ \forall k\in\mathcal{A}\}$.
            \ENDIF
        \ENDIF
    \ENDWHILE
    \STATE $T\leftarrow T+1$; $\mathcal{A}_T\leftarrow\mathcal{A}$.
\ENDWHILE
\RETURN $T$, $\{\mathcal{A}_t\}_{t=1}^{T}$, and the associated PA positions and powers.
\end{algorithmic}
\end{algorithm}

\subsection{Complexity and Convergence}
\label{subsec:multicell_properties}

We first analyze the computational complexity of the proposed algorithm. Constructing the weighted compatibility graph requires solving $\binom{K}{2}$ two-cell problems. Since the complexity of each two-cell solution is independent of $K$, the overall preprocessing complexity scales as $O(K^2)$. For a reuse group containing $n$ cells, each objective evaluation in the element-wise PA optimization requires solving an $n$-dimensional linear system, with complexity $O(n^3)$. Since each iteration updates $n$ PA positions through a $Q$-point one-dimensional search, the complexity of the group-wise refinement is $O(I_{\max}Qn^4)$, where $I_{\max}$ denotes the maximum number of iterations. The stepwise allocation examines at most $O(K^2)$ candidate admissions, which yields a conservative overall complexity of $O(I_{\max}QK^6)$. In practice, the compatibility graph reduces the number of candidate evaluations, and the reuse groups are typically much smaller than $K$.

Regarding convergence, each element-wise update cannot increase the required peak transmit power because the previous PA position is included in the search set. Since the objective is nonnegative, the resulting objective sequence converges. Moreover, the stepwise RB allocation terminates after assigning all $K$ cells, and each admitted group satisfies the SINR and transmit-power constraints. Nevertheless, due to the nonconvex PA-placement problem and greedy group construction, the proposed algorithm does not guarantee the globally minimum number of RBs.

\begin{figure}[!t]
\centering
    \subfigure[$R_{\rm{th}}=3$ bps/Hz.]
    {
        \includegraphics[width=0.45\textwidth]{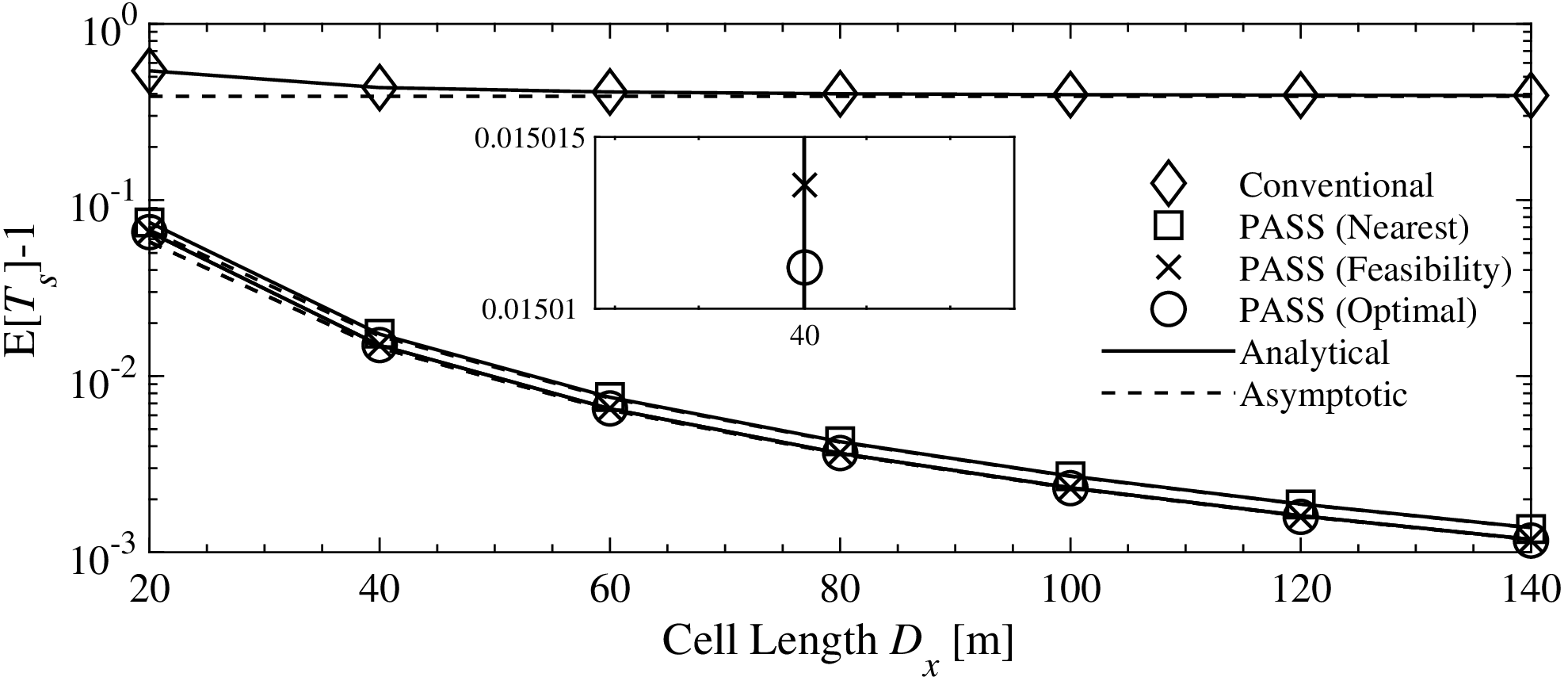}
	   \label{fig:Figure1a}
    }
    \subfigure[$D_x=80$ m.]
    {
        \includegraphics[width=0.45\textwidth]{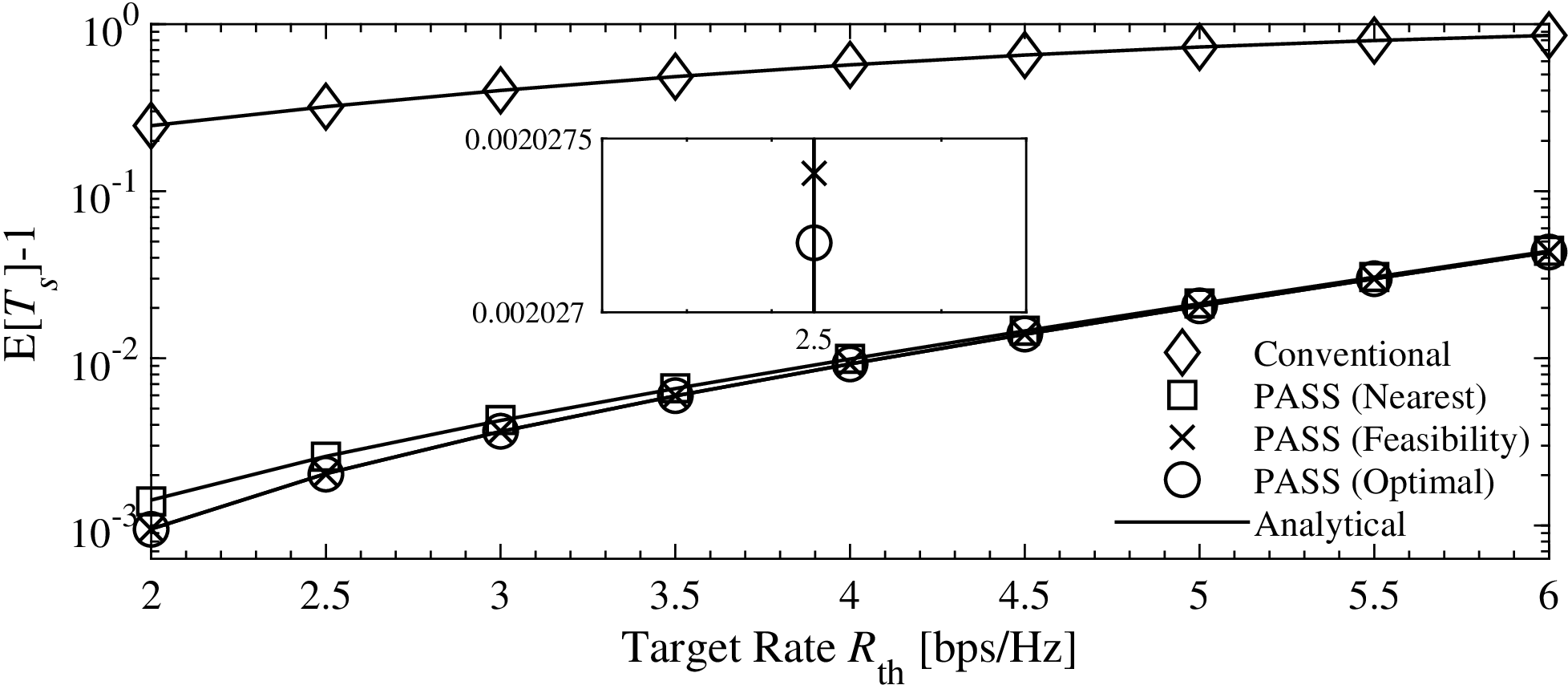}
	   \label{fig:Figure1b}
    }
\caption{Excess RB requirement for the 1D two-cell case.}
\label{fig:Figure1}
\vspace{-15pt}
\end{figure}

\section{Numerical Results}
\label{sec:numerical_results}
In this section, we evaluate the performance of multi-cell PASS in terms of RB requirements and compare it with a conventional fixed-antenna system, where each antenna is deployed at the cell center. Unless otherwise stated, the carrier frequency is $f_{\rm{c}}=28$ GHz, the noise power is $\sigma^2=-90$ dBm, and the waveguide height is $H=3$ m. The service area is divided into $N_{\rm row}$ rows and $N_{\rm col}$ columns, which yields $K=N_{\rm row}N_{\rm col}$ rectangular cells of equal size. Each cell has dimensions $D_x=80$ m and $D_y=20$ m and contains one uniformly distributed user. Each one-dimensional PA update searches a uniform grid of $Q = 2001$ points. We consider Gaussian signaling with interference treated as noise, where the SINR target $\gamma$ corresponds to a target rate of $R_{\rm th}=\log_2(1+\gamma)$. To guarantee that every user can be served on an individual RB, we set the transmit-power budget according to the worst-case coverage requirement of the conventional fixed-antenna system, i.e., $P_{\max}=\frac{\gamma\sigma^2(H^2+{D_x^2/4+D_y^2/4})}{\eta}
$. The power budget is updated accordingly when $D_x$ or $R_{\rm th}$ varies. 

\subsection{One-Dimensional Two-Cell Case}

We first consider the 1D two-cell case, where $D_y=0$ and both the waveguides and users are distributed along a line. As established in Section~\ref{subsec:average_rb}, the average RB requirement of PASS approaches one as $D_x\to\infty$. To better illustrate this asymptotic behavior, {\figurename} \ref{fig:Figure1} presents the excess RB requirement $\mathbb{E}[T_s]-1$ instead of $\mathbb{E}[T_s]$ for $s\in\{{\rm N},{\rm I},{\rm P},{\rm C}\}$.

{\figurename} {\ref{fig:Figure1a}} compares the conventional fixed-antenna system with PASS under nearest-PA, feasibility-optimal, and power-optimal PA placement. The analytical and asymptotic results for the conventional system are numerically evaluated using \eqref{eq:ETC_analy} and \eqref{eq:ETC_asymp}, respectively. For PASS, the analytical results are obtained from \eqref{eq:ET_reference}, while the asymptotic results follow from \eqref{eq:ETN_asymp}--\eqref{eq:ETP_asymp}. The analytical results closely match the simulations for all four schemes. In particular, although \eqref{eq:ET_reference} provides approximations for the feasibility-optimal and power-optimal strategies, which become asymptotically exact according to Lemma~\ref{lemma:average_rb}, the approximations remain highly accurate even for moderate values of $D_x$. 

As shown in {\figurename} \ref{fig:Figure1a}, all three PASS strategies significantly reduce the excess RB requirement compared with the conventional fixed-antenna system. This improvement arises from the flexibility of PA placement, which allows the desired-link power requirement and ICI to be jointly reduced. Furthermore, the feasibility-optimal and power-optimal strategies achieve nearly identical performance and both outperform nearest-PA placement, although the difference is relatively small. This indicates that nearest-PA placement already provides an effective low-complexity alternative for the considered two-cell system. As $D_x$ increases, the excess RB requirements for all four schemes decrease, primarily because larger cell dimensions reduce the likelihood of strong ICI. The excess RB requirements of all three PASS strategies approach zero, whereas that of the conventional system approaches a strictly positive constant, as characterized by \eqref{eq:ETC_asymp}. Moreover, the feasibility-optimal and power-optimal strategies exhibit smaller asymptotic coefficients than nearest-PA placement, while sharing the same decay order. These observations validate the asymptotic characterization and the insights presented in Remark~\ref{remark: convergence}. 

{\figurename} \ref{fig:Figure1b} further compares the four schemes with respect to the target rate $R_{\rm th}$. As expected, the excess RB requirement increases with $R_{\rm th}$, since a higher SINR target makes simultaneous transmission more difficult. Nevertheless, all three PASS strategies consistently outperform the conventional fixed-antenna system. The feasibility-optimal and power-optimal strategies provide additional gains over nearest-PA placement, particularly at relatively low target rates. As $R_{\rm th}$ increases, the performance differences among the three PASS strategies become less pronounced, which further demonstrates the effectiveness of nearest-PA placement.

\begin{figure}[!t]
\centering
    \subfigure[$R_{\rm{th}}=3$ bps/Hz.]
    {
        \includegraphics[width=0.45\textwidth]{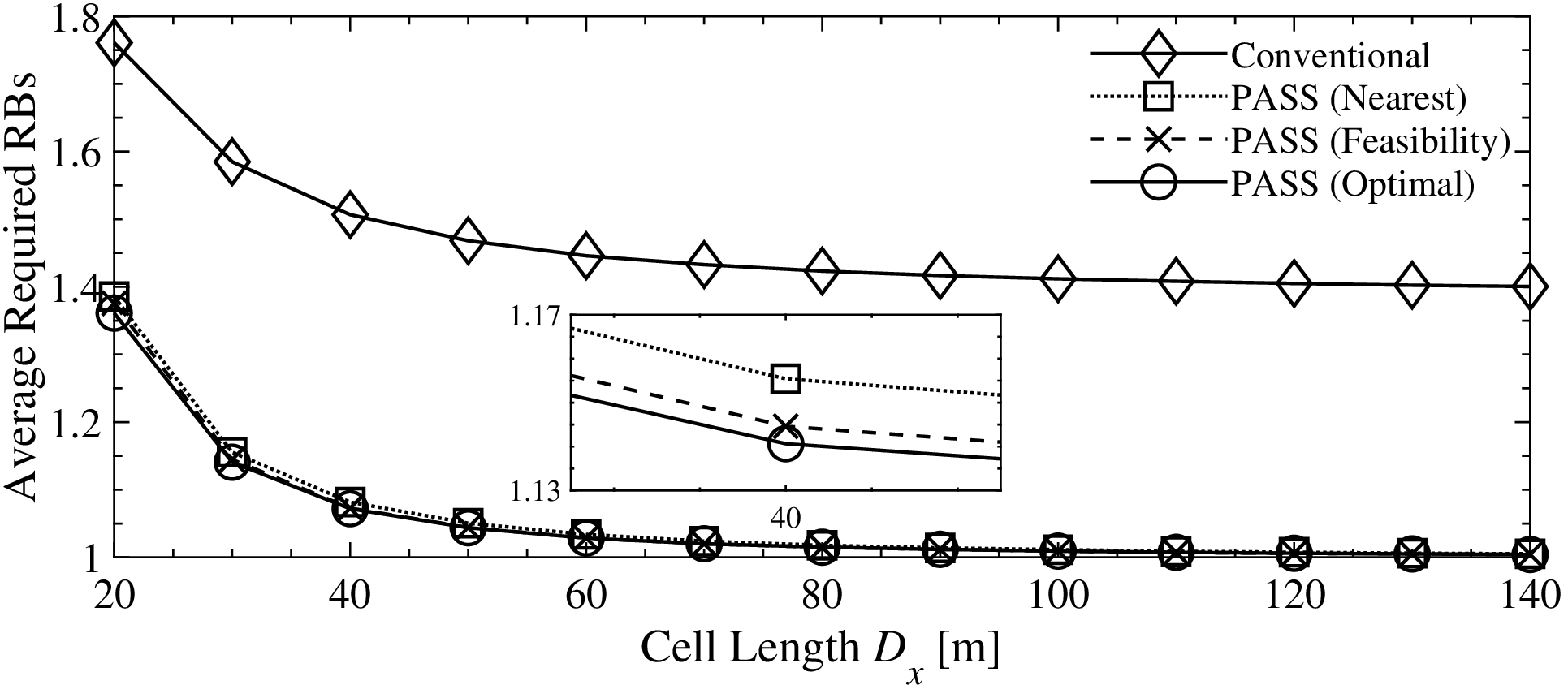}
	   \label{fig:Figure2a}
    }
    \subfigure[$D_x=80$ m.]
    {
        \includegraphics[width=0.45\textwidth]{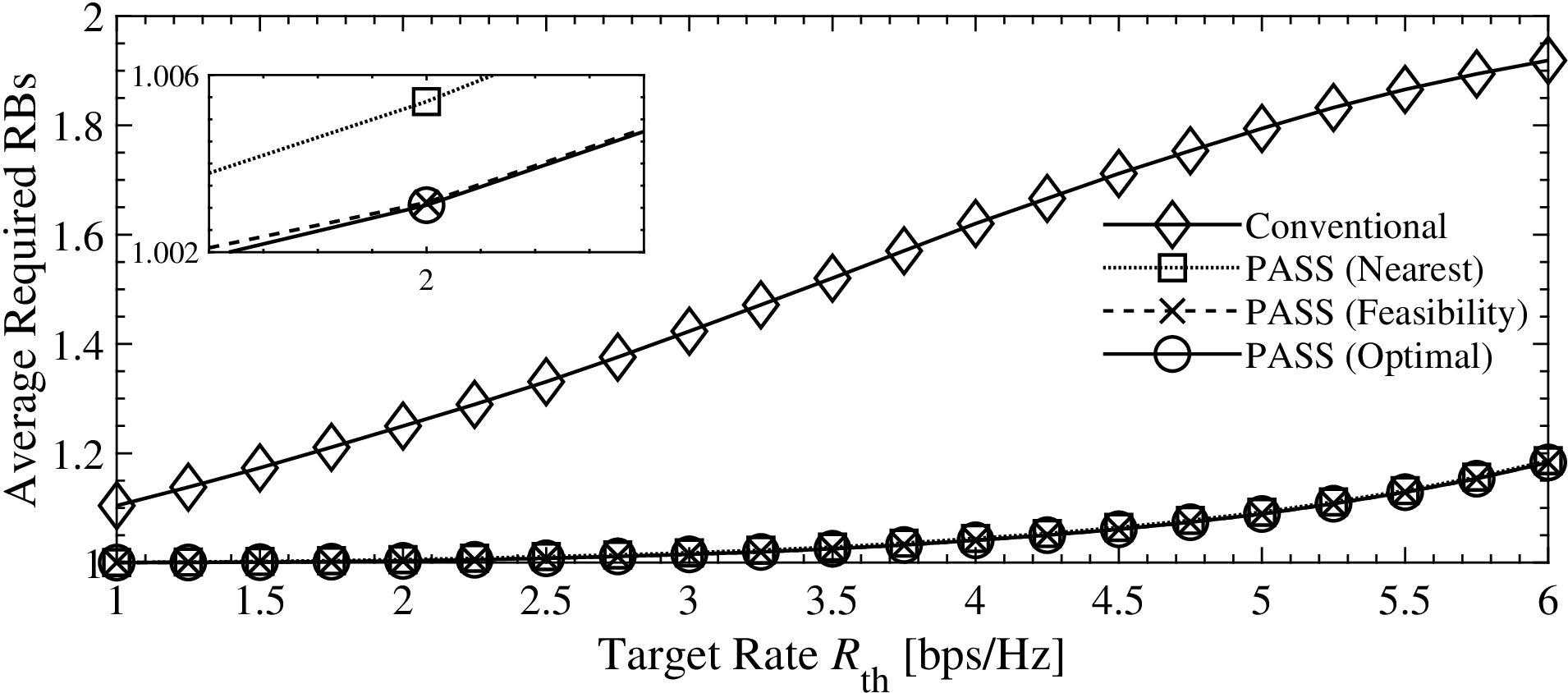}
	   \label{fig:Figure2b}
    }
\caption{Average RB requirement for the 2D two-cell case.}
\label{fig:Figure2}
\vspace{-10pt}
\end{figure}

\subsection{Two-Dimensional Two-Cell Case}

We next investigate the general two-dimensional (2D) two-cell case. For a more comprehensive comparison, in addition to the power-optimal PA placement developed in Section \ref{subsec:two_cell_solution} and nearest-PA placement, we consider feasibility-optimal PA placement. This strategy is obtained by minimizing $\frac{a(x)}{b(x)}$ and $\frac{e(y)}{c(y)}$ in \eqref{eq:two-cell_feasibility} following the approach in Section~\ref{subsec:feasibility_optimal}.

{\figurename} \ref{fig:Figure2a} illustrates the average RB requirement versus the cell length $D_x$. All three PASS strategies significantly outperform the conventional fixed-antenna system, and their average RB requirements decrease toward one as $D_x$ increases. This observation is consistent with the asymptotic insights obtained for the 1D two-cell case, although the corresponding analytical expressions do not directly apply to the general 2D geometry. Moreover, the feasibility-optimal and power-optimal strategies achieve slightly better performance than nearest-PA placement, while all three strategies exhibit increasingly similar RB requirements for larger $D_x$. This again suggests that nearest-PA placement provides an attractive performance-complexity tradeoff.

{\figurename} \ref{fig:Figure2b} presents the average RB requirement versus $R_{\rm th}$. The RB requirements of all four schemes increase with $R_{\rm th}$, but PASS consistently outperforms the conventional fixed-antenna system throughout the considered range. In particular, at high target rates, the conventional system requires nearly two RBs on average, which indicates that simultaneous transmission is feasible for only a small fraction of user deployments. In contrast, the three PASS strategies require considerably fewer RBs, indicating that flexible PA placement can preserve a high probability of simultaneous transmission even under stringent SINR requirements. The feasibility-optimal and power-optimal strategies again exhibit nearly identical performance and provide moderate improvements over nearest-PA placement.

\begin{figure}[!t]
\centering
    \subfigure[$R_{\rm{th}}=3$ bps/Hz.]
    {
        \includegraphics[width=0.45\textwidth]{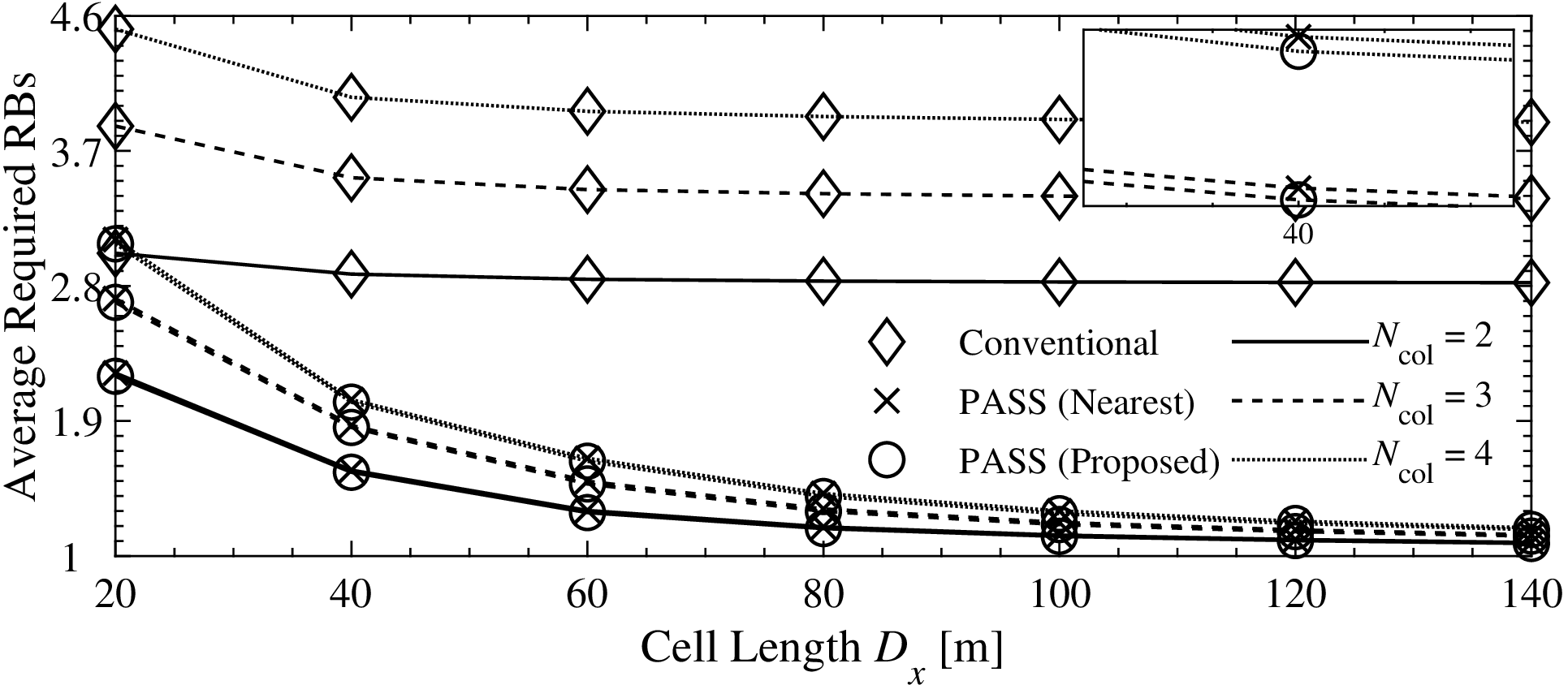}
	   \label{fig:Figure3a}
    }
    \subfigure[$D_x=80$ m.]
    {
        \includegraphics[width=0.45\textwidth]{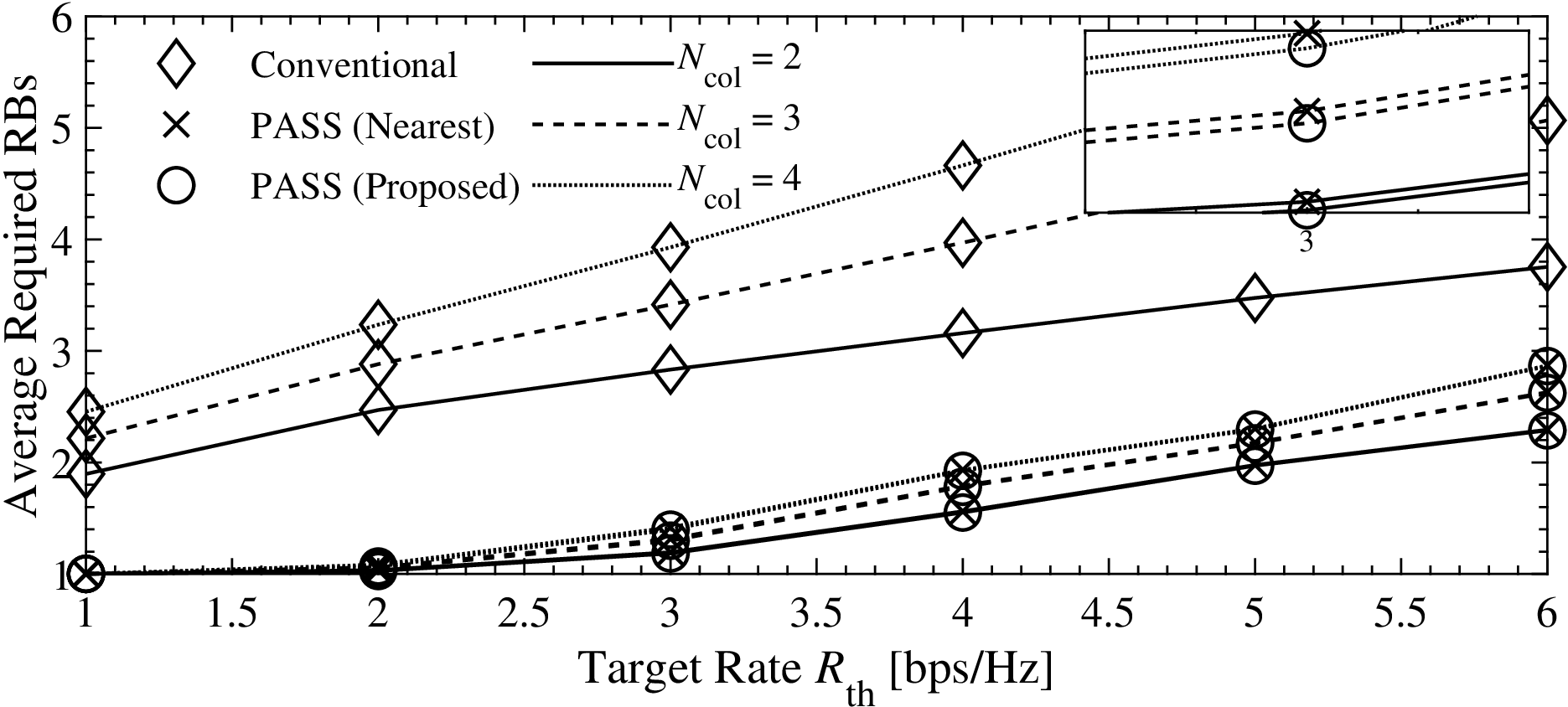}
	   \label{fig:Figure3b}
    }
\caption{Average RB requirement for the multi-cell case.}
\label{fig:Figure3}
\vspace{-10pt}
\end{figure}

\subsection{Multi-Cell Case}

We finally consider the general multi-cell case with $N_{\rm row}=2$. We compare three schemes: the conventional fixed-antenna system, nearest-PA placement, and the proposed joint RB allocation and PA-placement method. The conventional and nearest-PA schemes employ graph-assisted RB allocation with fixed antenna or PA positions, whereas the proposed method additionally optimizes PA positions through element-wise alternating optimization. 

{\figurename} \ref{fig:Figure3a} compares the average RB requirements of the three schemes versus $D_x$ for different numbers of cells. PASS significantly outperforms the conventional fixed-antenna system across all considered network sizes. In particular, as $D_x$ increases, the average RB requirements of PASS approach one, whereas the conventional system continues to require substantially more RBs. This extends the performance advantage observed in the two-cell cases to networks with multiple interfering cells. Moreover, the average RB requirement increases with $N_{\rm col}$, since more cells must be accommodated through spatial reuse or additional RB allocation. The proposed method achieves slightly lower RB requirements than nearest-PA placement, although their performances remain close. This suggests that graph-assisted RB allocation already captures a substantial portion of the available spatial-reuse gain, while PA-position refinement provides further improvements. {\figurename} \ref{fig:Figure3b} shows the average RB requirement versus $R_{\rm th}$. As the target rate increases, more RBs are required to satisfy the users' SINR constraints, and the increase becomes more pronounced as the number of cells grows. Nevertheless, both PASS schemes require much fewer RBs than the conventional fixed-antenna system. The proposed method slightly outperforms nearest-PA placement, although their average RB requirements remain close. These results demonstrate the effectiveness of PASS-assisted spatial reuse in multi-cell networks, particularly when the number of users or the target rate is large.

\begin{figure}[!t]
\centering
    \subfigure[$R_{\rm{th}}=3$ bps/Hz.]
    {
        \includegraphics[height=0.17\textwidth]{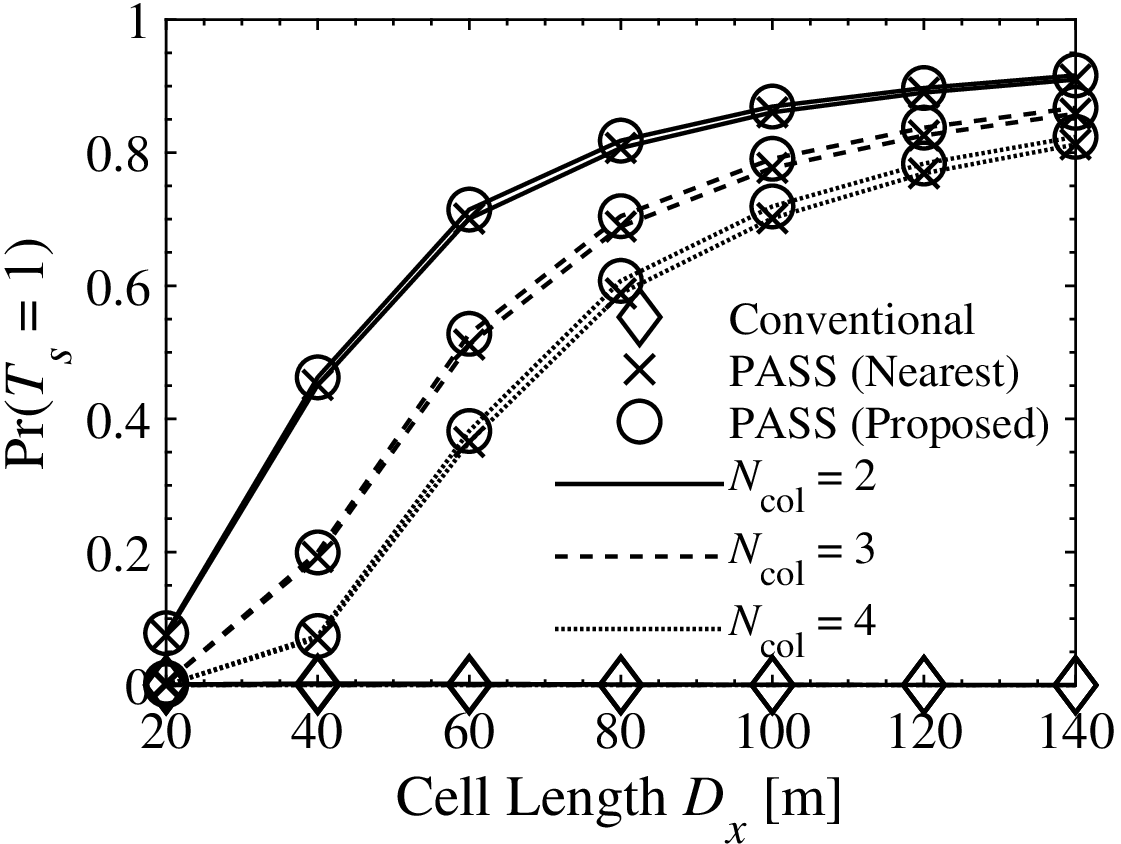}
	   \label{fig:Figure4a}
    }
    \subfigure[$D_x=80$ m.]
    {
        \includegraphics[height=0.17\textwidth]{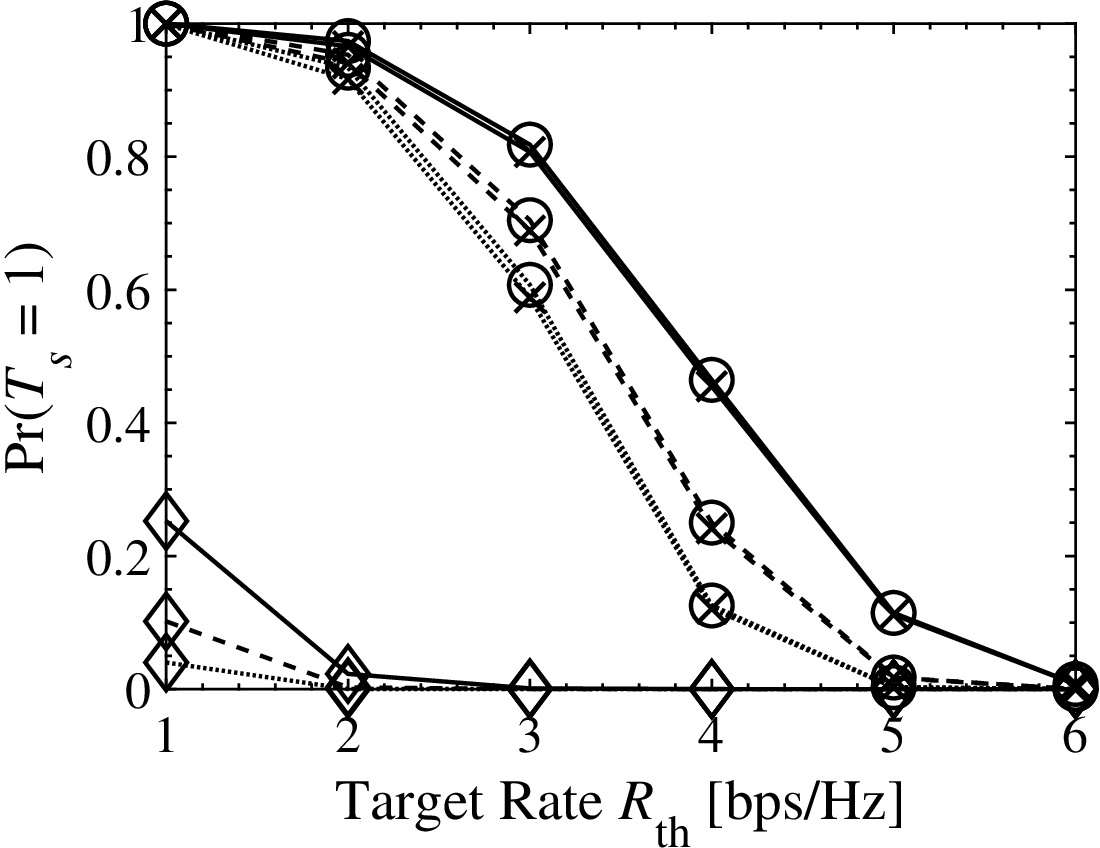}
	   \label{fig:Figure4b}
    }
\caption{Single-RB probability for the multi-cell case.}
\label{fig:Figure4}
\vspace{-10pt}
\end{figure}

To further examine the spatial-reuse capability, {\figurename} \ref{fig:Figure4} presents the probability $\Pr(T_s=1)$ that all $K$ cells can be served simultaneously using a single RB. As shown in {\figurename} \ref{fig:Figure4a}, this probability increases with $D_x$ for both PASS schemes, whereas it remains close to zero for the conventional fixed-antenna system over the considered configurations. Notably, PASS yields a high one-RB reuse probability even for networks with six or eight cells when $D_x$ is sufficiently large. {\figurename} \ref{fig:Figure4b} shows that the one-RB reuse probability decreases with $R_{\rm th}$. Nevertheless, PASS achieves a substantially higher probability than the conventional system, especially at moderate target rates. These observations demonstrate that PASS enables more effective spatial reuse through flexible antenna placement and interference-aware RB allocation.

\section{Conclusion}
\label{sec:conclusion}

In this paper, we investigated the potential of PASS to enhance spatial reuse in multi-cell networks. Both analytical and numerical results demonstrated that PASS substantially reduced the required RBs compared with conventional fixed-antenna systems. For the two-cell case, our analysis established that nearest-PA placement achieved the same asymptotic decay order as power-optimal placement for the excess RB requirement. For the general multi-cell case, numerical results further showed that nearest-PA placement combined with RB allocation achieved nearly the same performance as the proposed joint design. These findings highlighted the potential of nearest-PA placement as an intuitive and low-complexity alternative for practical multi-cell PASS deployment. 

From the perspective of 3GPP NR, the enhanced spatial reuse offered by PASS could enable more neighboring cells to simultaneously utilize the same physical resource blocks (PRBs), thus improving spectrum utilization and reducing the need for orthogonal resource allocation across cells. Future work will investigate how PA placement can simplify inter-cell coordination and resource scheduling in practical NR networks.

\begin{appendix}
\subsection{Proof of Lemma \ref{prop:symmetric_power_optimum}}\label{app:symmetric_power}
\begin{IEEEproof}
We first show that it suffices to consider $0\leq x_1,x_2\leq x_{\rm I}$. For any $x_i>x_{\rm I}$, replacing $x_i$ with $x_{\rm I}$ reduces both $A(x_i)$ and $r(x_i)$. From \eqref{eq:two_powers_norm_general}, this replacement strictly reduces both required transmit powers. Hence, any power-optimal solution must satisfy $0\leq x_1,x_2\leq x_{\rm I}$.

Define $L(x)\triangleq
\frac{1}{A(x)}+\frac{\gamma}{B(x)}
=\frac{1+\gamma r(x)}{A(x)}$. Its derivative satisfies $L'(x)
=-\frac{2x}{A^2(x)}
-\frac{2\gamma(d+x)}{B^2(x)}
<0$, so $L(x)$ is strictly decreasing for $x\geq0$. From \eqref{eq:two_powers_norm_general}, the ratio of the required transmit powers is $\frac{p_1}{p_2} =\frac{L(x_2)}{L(x_1)}$. Suppose $x_1>x_2$ without loss of generality. Since $L(x)$ is strictly decreasing, we have $p_1>p_2$, and the peak transmit power is $p_1$. Moreover, $\frac{\partial (p_1/p_0)}
{\partial r(x_2)}
=
\frac{
\gamma A(x_1)[1+\gamma r(x_1)]
}{
[1-\gamma^2r(x_1)r(x_2)]^2
}
>0$. Since $r(x_2)$ strictly decreases as $x_2$ increases toward $x_1\leq x_{\rm I}$, the peak transmit power $p_1$ strictly decreases. The relation $p_1>p_2$ holds as long as $x_2<x_1$. Hence, any placement with $x_1>x_2$ can be strictly improved by increasing $x_2$ toward $x_1$. The case $x_2>x_1$ follows by symmetry.

Therefore, the globally power-optimal PA placement satisfies $x_1^\star=x_2^\star$, which yields $p_1^\star=p_2^\star$.
\end{IEEEproof}
\subsection{Proof of Lemma~\ref{lemma:monotone}}
\label{app:optimal_displacement}
\begin{IEEEproof}
Differentiating $J(x,d)$ with respect to $x$ yields
$\frac{\partial J}{\partial x}
=\frac{2\gamma(d+x)}{B^2(x)}-\frac{2x}{A^2(x)}$.
Define $F(x)\triangleq\frac{xB^2(x)}{(d+x)A^2(x)}$. The stationary condition $\frac{\partial J}{\partial x}=0$ is equivalent to $F(x)=\gamma$. For $0<x<x_{\rm I}$, we have $\frac{F'(x)}{F(x)}
=
\frac{d}{x(d+x)}
+
\frac{4d[1-x(d+x)]}{A(x)B(x)}
>0$, where the inequality follows from $x(d+x)<1$ for $x<x_{\rm I}$. Thus, $F(x)$ is strictly increasing over $(0,x_{\rm I})$. Moreover, $F(0)=0$ and $F(x_{\rm I})=1/x_{\rm I}^2$. Since $d>d_{\rm F}$ implies $\gamma x_{\rm I}^2<1$, we obtain $F(0)<\gamma<F(x_{\rm I})$. Therefore, $F(x)=\gamma$ admits a unique solution $x_{\rm P}\in(0,x_{\rm I})$. Furthermore, $\frac{\partial J}{\partial x}
=\frac{2(d+x)}{B^2(x)}[\gamma-F(x)]$. Since $F(x)$ is strictly increasing, $J(x,d)$ strictly increases for $x<x_{\rm P}$ and strictly decreases for $x>x_{\rm P}$. Hence, $x_{\rm P}$ is the unique global maximizer of $J(x,d)$ over $[0,x_{\rm I}]$. Finally, since $r(x_{\rm I})=x_{\rm I}^2$ and $1-\gamma x_{\rm I}^2>0$ for $d>d_{\rm F}$, we have $J(x_{\rm P},d)
>
J(x_{\rm I},d)
=
\frac{1-\gamma x_{\rm I}^2}{1+x_{\rm I}^2}
>0$. Together with $J(x_{\rm P},d)=\frac{1-\gamma r(x_{\rm P})}{A(x_{\rm P})}$ and $A(x_{\rm P})>0$, this yields $1-\gamma r(x_{\rm P})>0$, which guarantees the existence of a finite positive transmit-power solution. 
\end{IEEEproof}
\subsection{Proof of Lemma~\ref{lemma_opt_power}}
\label{app:power_threshold}
\begin{IEEEproof}
By Lemma~\ref{lemma:monotone}, $x_{\rm P}(d)$ is the unique interior maximizer of $J(x,d)$ and satisfies $\frac{\partial J}{\partial x}=0$. Since $F'(x_{\rm P})>0$, the implicit function theorem ensures that $x_{\rm P}(d)$ is continuously differentiable. Consequently, $\frac{{\rm d}\overline{J}(d)}{{\rm d}d}
=
\frac{\partial J(x_{\rm P},d)}{\partial d}
+
\frac{\partial J(x_{\rm P},d)}{\partial x}
\frac{{\rm d}x_{\rm P}}{{\rm d}d}
=
\frac{2\gamma(d+x_{\rm P})}
{[1+(d+x_{\rm P})^2]^2}
>0$. Thus, $\overline{J}(d)$ is continuous and strictly increasing for $d>d_{\rm F}$.

We next characterize its limiting values. At $d=d_{\rm F}$, we have $r(x)\geq r(x_{\rm I})=x_{\rm I}^2=\frac{1}{\gamma}$ for all $x\geq0$. Hence, $J(x,d_{\rm F})
=\frac{1-\gamma r(x)}{A(x)}
\leq0$, with equality at $x=x_{\rm I}$. Therefore, the maximum value of $J(x,d_{\rm F})$ over $[0,x_{\rm I}]$ is exactly zero, i.e., $\max_{0\leq x\leq x_{\rm I}(d_{\rm F})}
J(x,d_{\rm F})=0$. Since $J(x,d)$ and $x_{\rm I}(d)$ are continuous in $d$, the maximum of $J(x,d)$ over $[0,x_{\rm I}(d)]$ is also continuous. It follows that
$\lim_{d\downarrow d_{\rm F}}\overline{J}(d)=0$.

At the other extreme, for any $x\geq0$, we have
$J(x,d)=\frac{1}{1+x^2}-\frac{\gamma}{1+(d+x)^2}<1$.
Thus, $\overline{J}(d)<1$. Meanwhile, the feasible choice $x=0$ yields $\overline{J}(d)
\geq J(0,d)=1-\frac{\gamma}{1+d^2}$. Combining these bounds gives $1-\frac{\gamma}{1+d^2}
\leq\overline{J}(d)<1$. Since $\lim_{d\to\infty}(1-\frac{\gamma}{1+d^2})=1$, the sandwich rule yields
$\lim_{d\to\infty}\overline{J}(d)=1$. Together with strict monotonicity, these limits establish $0<\overline{J}(d)<1$ for all $d>d_{\rm F}$. 
\end{IEEEproof}

\subsection{Proof of Lemma~\ref{lemma:average_rb}}
\label{app:average_rb}
\begin{IEEEproof}
Recall that $\bar{x}_1=\frac{D_x+x_1^{\rm U}}{H}$, $\bar{x}_2=\frac{D_x-x_2^{\rm U}}{H}$, and $d=\frac{x_2^{\rm U}-x_1^{\rm U}}{H}$, where $x_1^{\rm U}\in[-D_x,0]$ and $x_2^{\rm U}\in[0,D_x]$. Thus, $\bar{x}_1=r-d+\frac{x_2^{\rm U}}{H}\geq r-d$ and $\bar{x}_2=r-d-\frac{x_1^{\rm U}}{H}\geq r-d$.

We first consider $d\leq r-1$. In this case, $\bar{x}_i\geq r-d\geq1>x_{\rm I}$ for $i\in\{1,2\}$. Thus, the feasibility-optimal PA placements are attainable. For $d>d_{\rm F}$, the power-optimal placements are also attainable since $x_{\rm P}<x_{\rm I}$. Hence, the previously derived reuse conditions apply directly.

For $d>r-1$, the available PA offsets may be insufficient to attain $x_{\rm I}$. Nevertheless, the feasibility-optimal offsets $x_i=\min\{x_{\rm I},\bar{x}_i\}$ satisfy $0\leq x_i\leq x_{\rm I}<d^{-1}$. Since $r(x)$ decreases over $[0,x_{\rm I}]$, we have $A(x_i)\leq1+d^{-2}$ and $r(x_i)\leq r(0)<d^{-2}$. For sufficiently large $r$, we have $d>r-1>\sqrt{\gamma}$. From \eqref{eq:two_powers_norm_general}, the required peak transmit power is thus bounded by
\begin{align*}
\frac{\max\{p_1,p_2\}}{p_0}
&\leq
\frac{(1+d^{-2})(1+\gamma d^{-2})}
{1-\gamma^2d^{-4}}
=\frac{1+d^{-2}}{1-\gamma d^{-2}}\\
&\leq
\frac{1+(r-1)^{-2}}
{1-\gamma(r-1)^{-2}}.
\end{align*}
The last upper bound approaches one as $r\to\infty$, whereas $\rho=1+\frac{r^2}{4}\to\infty$. Hence, feasibility-optimal placement supports one-RB reuse for sufficiently large $r$. The same conclusion holds for nearest-PA placement, whose required transmit power is also bounded by the above expression, and for power-optimal placement, which minimizes the required peak transmit power.

As derived in Sections~\ref{sec:required_rb_Nearest-PA Placement} and \ref{sec:required_rb_Feasibility-Optimal PA Placement}, $d_{\rm N}$ and $d_{\rm I}$ remain bounded as $r\to\infty$. Together with $d_{\rm F}<d_{\rm P}<d_{\rm I}$ in \eqref{eq:reuse_threshold_order}, this implies $d_s<r-1$ for $s\in\{{\rm N},{\rm I},{\rm P}\}$ when $r$ is sufficiently large. Consequently, the average number of required RBs is
\begin{align*}
\mathbb{E}[T_s]
&=\Pr(d>r-1)+\Pr(d_s\leq d\leq r-1)
+2\Pr(d<d_s)\\
&=\Pr(d\geq d_s)+2\Pr(d<d_s)\\
&=1-\Pr(d<d_s)+2\Pr(d<d_s)\\
&=1+\Pr(d<d_s)
=1+\Pr(D<Hd_s)\\
&=1+F_D(Hd_s).
\end{align*}
Since $d_s$ remains bounded as $r\to\infty$, we have $Hd_s<D_x=Hr$ for sufficiently large $r$. Therefore, using the CDF in \eqref{eq:D_cdf}, we obtain $1+F_D(Hd_s)=
1+\frac{H^2d_s^2}{2D_x^2}
=
1+\frac{d_s^2}{2r^2}$.
\end{IEEEproof}

\end{appendix}
\bibliographystyle{IEEEtran}
\bibliography{mybib}

@article{wu2019towards,
  title={Towards smart and reconfigurable environment: Intelligent reflecting surface aided wireless network},
  author={Wu, Qingqing and Zhang, Rui},
  journal={IEEE communications magazine},
  volume={58},
  number={1},
  pages={106--112},
  year={2019},
  publisher={IEEE}
}

@article{ouyang2024primer,
  title={A primer on near-field communications for next-generation multiple access},
  author={Ouyang, Chongjun and Wang, Zhaolin and Chen, Yan and Mu, Xidong and Zhu, Peiying},
  journal={Proc. {IEEE}},
  volume={112},
		number={9},
		pages={1527--1565},
		year={2024},
		month=sep,
		publisher={IEEE}
}

@article{ding2024flexible,
  title={Flexible-Antenna Systems: A Pinching-Antenna Perspective},
  author={Ding, Zhiguo and Schober, Robert and Poor, H Vincent},
  journal={IEEE Trans. Commun.},
  volume={73},
  number={10},
  pages={9236--9253},
  year={2025},
  month=oct,
  publisher={IEEE}
}

@article{suzuki2022pinching,
  title={Pinching Antenna: Using a Dielectric Waveguide as an Antenna},
  author={Fukuda, Atsushi and Yamamoto, Hiroto and Okazaki, Hiroshi and Suzuki, Yasunori and Kawai, Kunihiro},
  journal={NTT DOCOMO Technical J.},
  volume={23},
  number={3},
  pages={5--12},
  year={2022},
  month=jan
}

@article{zhu2024movable,
  title={Movable antennas for wireless communication: Opportunities and challenges},
  author={Zhu, Lipeng and Ma, Wenyan and Zhang, Rui},
  journal={IEEE Commun. Mag.},
  volume={62},
  number={6},
  pages={114--120},
  year={2024},
  month=jun,
  publisher={IEEE}
}

@article{wong2020fluid,
  title={Fluid antenna systems},
  author={Wong, Kai-Kit and Shojaeifard, Arman and Tong, Kin-Fai and Zhang, Yangyang},
  journal={IEEE Trans. Wireless Commun.},
  volume={20},
  number={3},
  pages={1950--1962},
  year={2021},
  month=mar,
  publisher={IEEE}
}

@article{liu2025pinching,
  title={Pinching antenna systems {(PASS)}: Architecture designs, opportunities, and outlook},
  author={Liu, Yuanwei and Wang, Zhaolin and Mu, Xidong and Ouyang, Chongjun and Xu, Xiaoxia and Ding, Zhiguo},
  journal={IEEE Commun. Mag.},
  volume={64},
  number={1},
  pages={190--196},
  year={2026},
  month=jan,
  publisher={IEEE}
}

@article{xu2026pinching,
  title={Pinching-antenna systems with in-waveguide attenuation: Performance analysis and algorithm design},
  author={Xu, Yanqing and Ding, Zhiguo and Schober, Robert and Chang, Tsung-Hui},
  journal={IEEE Trans. Wireless Commun.},
  volume={25},
  pages={14564--14580},
  year={2026},
  publisher={IEEE}
}

@book{yeh2008essence,
  title={The Essence of Dielectric Waveguides},
  author={Yeh, Cavour and Shimabukuro, Fred I},
  year={2008},
  publisher={New York, NY, USA: Springer}
}

@article{liu2025pinchingtutorial,
  title={Pinching-antenna systems ({PASS}): A tutorial},
  author={Liu, Yuanwei and Jiang, Hao and Xu, Xiaoxia and Wang, Zhaolin and Guo, Jia and Ouyang, Chongjun and Mu, Xidong and Ding, Zhiguo and Nallanathan, Arumugam and Karagiannidis, George K and others},
  journal={IEEE Trans. Commun.},
  volume={74},
  pages={4881--4918},
  year={2026}
}

@article{liu2025pinchingsurvey,
  title={A Survey of Pinching-Antenna Systems ({PASS})},
  author={Liu, Yuanwei and Jiang, Hao and Gan, Xu and Xu, Xiaoxia and Guo, Jia and Wang, Zhaolin and Ouyang, Chongjun and Mu, Xidong and Ding, Zhiguo and Nallanathan, Arumugam and Dobre, Octavia A and Karagiannidis, George K and Schober, Robert},
  journal={IEEE Commun. Surveys Tuts.},
  volume={28},
  pages={7250--7280},
  year={2026},
  publisher={IEEE}
}

@article{lu2026survey,
  title={A Survey on Pinching-Antenna Systems: Optimization Techniques, Intelligent Designs, and Multifunctional Applications},
  author={Lu, Yang and Mao, Weihao and Xie, Xinke and Liao, Shuhai and Xu, Yanqing and Lin, Yushen and Zhang, Ruichen and Ai, Bo and Wang, Xianbin and Ding, Zhiguo},
  journal={IEEE Commun. Surveys Tuts.},
  volume={28},
  pages={6592--6627},
  year={2026},
  publisher={IEEE}
}

@article{shlezinger2021dynamic,
  title={Dynamic metasurface antennas for {6G} extreme massive {MIMO} communications},
  author={Shlezinger, Nir and Alexandropoulos, George C and Imani, Mohammadreza F and Eldar, Yonina C and Smith, David R},
  journal={IEEE Wireless Commun.},
  volume={28},
  number={2},
  pages={106--113},
  year={2021},
  month=apr,
  publisher={IEEE}
}

@article{ouyang2025uplink,
  title={Uplink and Downlink Communications in Segmented Waveguide-Enabled Pinching-Antenna Systems ({SWANs})},
  author={Ouyang, Chongjun and Jiang, Hao and Wang, Zhaolin and Liu, Yuanwei and Ding, Zhiguo},
  journal={IEEE Trans. Commun.},
  volume={74},
  pages={3688--3703},
  year={2026},
  publisher={IEEE}
}

@article{xu2026pinching2,
  title={Pinching-antenna system design with LoS blockage: Does in-waveguide attenuation matter?},
  author={Xu, Yanqing and Ding, Zhiguo and Dobre, Octavia A and Chang, Tsung-Hui},
  journal={IEEE Trans. Wireless Commun.},
  volume={25},
  pages={17322--17337},
  year={2026},
  publisher={IEEE}
}

@book{cox1998using,
  title={Using Algebraic Geometry},
  author={Cox, David A and Little, John and O'shea, Donal},
  edition={2},
  year={2005},
  publisher={New York, NY, USA: Springer}
}

@article{ding2026toward,
  title={Toward a Quiet Wireless World: Multi-Cell Pinching-Antenna Transmission},
  author={Ding, Zhiguo},
  journal={IEEE Wireless Commun. Lett.},
  volume={15},
  pages={3791--3795},
  year={2026}
}

@article{ding2026characterizing,
  title={Characterizing Multi-Cell Pinching-Antenna Transmission: Revealing the Other Side of the Coin},
  author={Ding, Zhiguo},
  journal={arXiv preprint arXiv:2609.09585},
  year={2026}
}

@article{sun2025stochastic,
  title={A stochastic geometric analysis on multi-cell pinching-antenna systems under blockage effect},
  author={Sun, Yanshi and Ding, Zhiguo and Karagiannidis, George K},
  journal={IEEE Wireless Commun. Lett.},
  volume={15},
  pages={1085--1089},
  year={2025},
  publisher={IEEE}
}

@article{zhu2025topological,
  title={Topological perspective of large-scale multi-cell deployment of excitable waveguide dielectrics},
  author={Zhu, Yongxu and Ding, Zhiguo and You, Xiaohu},
  journal={IEEE Wireless Commun. Lett.},
  volume={15},
  pages={151--155},
  year={2025},
  publisher={IEEE}
}

@article{chen2026pinching,
  title={Pinching-Antenna Enabled Multicell Wireless Systems},
  author={Chen, Yunshu and Xue, Qing and Hua, Meng and Zhou, Bingpeng and Ma, Shaodan},
  journal={IEEE Trans. Wireless Commun.},
  volume={15},
  pages={24226--24240},
  year={2026},
  publisher={IEEE}
}

@article{li2026pinching,
  title={Pinching antenna systems ({PASS}) for cell-free communications},
  author={Li, Haochen and Lei, Jiayi and Zeng, Cheng and Hu, Zhaoming and Dong, Chao and Liu, Yuanwei},
  journal={IEEE Trans. Veh. Tech.},
  year={Early Access, 2026},
  publisher={IEEE}
}

@article{jiang2026pinching,
  title={Pinching Antennas-Enabled Cell-Free Massive {MIMO}: Two-Stage {AP--UE} Association and Pinching Beamforming Optimization},
  author={Jiang, Yancheng and Zhang, Jiayi and Liu, Zhilong and Ouyang, Chongjun and Zheng, Jiakang and Liu, Yuanwei and Nallanathan, Arumugam and Ai, Bo},
  journal={IEEE Trans. Veh. Tech.},
  year={Early Access, 2026},
  publisher={IEEE}
}
\end{document}